\global\def\draftcontrol{0}

   \def\versionno{ kt correlators -- draft   }

\catcode`\@=11

\expandafter\ifx\csname draftcontrol\endcsname\relax\global\def\draftcontrol{0}
\fi

{\count255=\time\divide\count255 by 60
\xdef\hourmin{\number\count255}
\multiply\count255 by-60\advance\count255 by\time
\xdef\hourmin{\hourmin:\ifnum\count255<10 0\fi\the\count255}}
\def\draftdate{\number\month/\number\day/\number\year\ \ \ \hourmin }

\newcommand\makepapertitle{\par
  \begingroup
    \renewcommand\thefootnote{\@fnsymbol\c@footnote}%
    \def\@makefnmark{\rlap{\@textsuperscript{\normalfont\@thefnmark}}}%
    \long\def\@makefntext##1{\parindent 1em\noindent
            \hb@xt@1.8em{%
                \hss\@textsuperscript{\normalfont\@thefnmark}}##1}%
     \newpage
     \global\@topnum\z@   
     \@makepapertitle
     \thispagestyle{empty}\@thanks
  \endgroup
  \setcounter{footnote}{0}%
  \global\let\thanks\relax
  \global\let\makepapertitle\relax
  \global\let\@makepapertitle\relax
  \global\let\@thanks\@empty
  \global\let\@author\@empty
  \global\let\@date\@empty
  \global\let\@title\@empty
  \global\let\title\relax
  \global\let\author\relax
  \global\let\date\relax
  \global\let\and\relax
  \def\version{\let\version\@version\@gobble}
}
\def\@makepapertitle{%
  \newpage
   \ifnum\draftcontrol=1 {}
   \version\versionno
   \vskip 3em%
   \else
   \hfill\hbox to 3cm {\parbox{5cm}{\@pubnum}\hss}%
   \vskip 3em%
   \fi
   \begin{center}%
   \let \footnote \thanks
     {\LARGE {\@title}}%
     \vskip 1.5em%
     {\normalsize
       \lineskip .5em%
       \begin{tabular}[t]{c}%
         \@author
       \end{tabular}\par}%
     \vskip 1.5em%
     {\@bstract}%
     \end{center}%
     \vskip 1.5em
     \@date%
   \par
}

\gdef\@pubnum{}
\def\pubnum#1{%
  \gdef\@pubnum{#1}}

\gdef\@bstract{}
\def\Abstract#1{%
  \gdef\@bstract{%
   \parbox{\textwidth-0pc}{%
   \centerline{\bf Abstract}\penalty1000
   \noindent
   \renewcommand\baselinestretch{1.0}
   {#1}}}
}

\def\ps@paper{\let\@mkboth\@gobbletwo%
     \ifnum\draftcontrol=1
        \def\@oddfoot{\hbox to \textwidth{\tiny \versionno \hfil\tiny\draftdate}%
        \hskip -\textwidth \hbox to \textwidth{\hfil\rm\thepage\hfil}}%
     \else\def\@oddfoot{\hbox to \textwidth{\hfil\rm\thepage\hfil}}
     \fi
     \let\@evenfoot\@oddfoot
}

\def\body{\clearpage
          \pagestyle{paper}
        }

\def\@version#1{\ifnum\draftcontrol=1
\typeout{}\typeout{#1}\typeout{}
\vskip3mm\centerline{\hbox{\fbox{\normalsize{\tt DRAFT -- #1 -- }
                   {\draftdate}}}}\vskip3mm
\fi}
\let\version\@version
\long\def\eqlabel#1{\ifnum\draftcontrol=1
                    \tag@false  
                    \tag*{(\theequation) \hbox to -0.2cm{\hspace{0cm}\small{#1}\hss}}
                    \refstepcounter{equation}
                    \edef\@currentlabel{\theequation}
                    \ltx@label{#1}          
                    \else
                    \label{#1}
                    \fi
                    }
\let\st@bibitem\@bibitem
\let\st@lbibitem\@lbibitem
\ifnum\draftcontrol=1
  \def\@bibitem#1{%
    \st@bibitem{#1}\a@@label{#1}\ignorespaces}
  \def\@lbibitem[#1]#2{%
    \st@lbibitem[#1]{#2}\a@@label{#2}\ignorespaces}
  \def\a@@label#1{%
    \gdef\a@lab{\smash{\normalfont\small#1}}
    \ifvmode
      \if@inlabel
        \global\setbox\@labels\hbox{%
          \llap{\a@lab\let\a@lab\relax
                \kern\@totalleftmargin\kern\marginparsep}%
          \box\@labels}%
      \fi
    \fi}
\fi

\documentclass[12pt,letterpaper]{article}

\usepackage{amsmath,amssymb,array,calc,rotating,epsfig,psfrag,cite,xspace}

\ifnum\draftcontrol=1
\fi

\renewcommand\baselinestretch{1.25}
\renewcommand\section{\@startsection {section}{1}{\z@}%
                                   {-3.5ex \@plus -1ex \@minus -.2ex}%
                                   {2.3ex \@plus.2ex}%
                                   {\normalfont\large\bfseries}}
\renewcommand\subsection{\@startsection{subsection}{2}{\z@}%
                                   {-3.25ex\@plus -1ex \@minus -.2ex}%
                                   {1.5ex \@plus .2ex}%
                                   {\normalfont\normalsize\bfseries}}
\renewcommand\subsubsection{\@startsection{subsubsection}{3}{\z@}%
                                   {-3.25ex\@plus -1ex \@minus -.2ex}%
                                   {1.5ex \@plus .2ex}%
                                   {\normalfont\normalsize\it}}
\renewcommand\paragraph{\@startsection{paragraph}{4}{\z@}%
                                   {-3.25ex\@plus -1ex \@minus -.2ex}%
                                   {1.5ex \@plus .2ex}%
                                   {\normalfont\normalsize\bf}}

\numberwithin{equation}{section}

\def\ie{{\it i.e.}}

\def\revise#1       {\raisebox{-0em}{\rule{3pt}{1em}}%
                     \marginpar{\raisebox{.5em}{\vrule width3pt\
                     \vrule width0pt height 0pt depth0.5em
                     \hbox to 0cm{\hspace{0cm}{%
                     \parbox[t]{4em}{\raggedright\footnotesize{#1}}}\hss}}}}

\newcommand\nxt[1]  {\\\fnxt#1}

\def\cala         {{\cal A}}

\def\caln         {{\cal N}}
\def\calo         {{\cal O}}

\def\del          {\partial}

\def\sqr#1#2{{\vcenter{\vbox{\hrule height.#2pt
 \hbox{\vrule width.#2pt height#1pt \kern#1pt
 \vrule width.#2pt}\hrule height.#2pt}}}}
\def\square{%
  \mathop{\mathchoice{\sqr{12}{15}}{\sqr{9}{12}}{\sqr{6.3}{9}}{\sqr{4.5}{9}}}}

\newcommand{\ft}[2]{{\textstyle{\frac{#1}{#2}}}}

\def\r{\rho}

\def\dd{\delta}

\def\vev#1{\langle #1 \rangle}

\def\k{\kappa}

\def\nb{\bar{n}}
\def\mb{\overline{m}}

\catcode`\@=12

\begin{document}


\title{Short distance properties of cascading gauge theories}

\pubnum{
WIS/11/06-AUG-DPP\\
UWO-TH/06/12}
\date{August 2006}

\author{Ofer Aharony$ ^1$, Alex Buchel$ ^{2,3,4}$ and
Amos Yarom$ ^{5}$\\[0.4cm]
\it $^1$Department of Particle Physics, Weizmann Institute of Science,\\
\it Rehovot 76100, Israel\\[0.2cm]
\it $^2$Perimeter Institute for Theoretical Physics\\
  \it Waterloo, Ontario N2J 2W9, Canada\\[0.2cm]
  \it $^3$Department of Applied Mathematics, University of Western Ontario\\
  \it London, Ontario N6A 5B7, Canada\\[0.2cm]
\it $^4$Albert Einstein Minerva Center, Weizmann Institute of Science,\\
\it Rehovot 76100, Israel\\[0.2 cm]
\it $^5$Department of Physics, Ben-Gurion University, Be'er Sheva 84105,
Israel\\
 }

\Abstract{We study the short distance (large momentum) properties of
correlation functions of cascading gauge theories by performing a
tree-level computation in their dual gravitational background. We
prove that these theories are holographically renormalizable; the
correlators have only analytic ultraviolet divergences, which may be
removed by appropriate local counterterms. We find that $n$-point
correlation functions of properly normalized operators have the
expected scaling in the semi-classical gravity (large $N$) limit:
they scale as $N_{eff}^{2-n}$ with $N_{eff}\propto \ln(k/\Lambda)$
where $k$ is a typical momentum. Our analysis thus confirms the
interpretation of the cascading gauge theories as renormalizable
four-dimensional quantum field theories with an effective number of
degrees of freedom which logarithmically increases with the energy.}


\makepapertitle

\body

\version\versionno

\section{Introduction and summary}

Cascading gauge theories were discovered in \cite{kn,KT,ks} (see
\cite{Herzog:2002ih,Strassler:2005qs} for reviews) by looking at the
decoupling limit (the near-horizon limit) of fractional D3-branes at
a conifold singularity. Since then various other
examples have also been studied, including
\cite{Gauntlett:2004zh,Gauntlett:2004yd,
Benvenuti:2004dy,Martelli:2004wu,Herzog:2004tr,Franco:2005fd,or1,or2}.
These theories are not standard local quantum field theories since
they do not approach a conformal field theory at high energies.
Therefore, one cannot use standard field theory techniques to
analyze them. When one introduces a finite high-energy cutoff at
some scale $M$, then at the cutoff scale these theories resemble an
$\caln=1$ supersymmetric $SU(K)\times SU(K+P)$ gauge theory with two
bifundamental and two anti-bifundamental chiral superfields and some
superpotential. When one flows down in energy from this cutoff one
of the gauge groups becomes strongly coupled and the theory seems to
undergo a series of Seiberg duality \cite{Seiberg:1994pq}
``cascades'', reducing the value of
$K$, and finally ending (when $K$ is a multiple of $P$)
with a confining theory at some low-energy
scale $\Lambda$ \cite{ks,Strassler:2005qs,Dymarsky:2005xt} (which is
related to
the $\caln=1$ supersymmetric pure $SU(P)$
Yang-Mills theory). However, the value
of $K$ increases with the high-energy (UV) cutoff as $K
\propto \ln(M/\Lambda)$, so
it seems that an increasing number of
degrees of freedom is needed to define the theory at higher
energies, and that the ultimate definition of the cascading theory
requires a theory with an infinite number of fields.

It is not known how to directly define a cascading gauge theory in
field theory terms\footnote{It may be possible to define it by a
limiting procedure, using an infinite-number-of-fields-limit of
well-defined theories which flow to the cascade at some energy
scale, as in the construction of \cite{Hollowood:2004ek}.}. The best
available definition of the cascading gauge theory is via its
holographic dual background\cite{kn,KT,ks}. This background can be
well-described by a semi-classical supergravity theory when the
dimensionless parameter of the cascading gauge theories, $g_s P$
(where $g_s$ is the string coupling in the dual background), is
large. However, the asymptotic region of this background is actually
well-described by supergravity for any value of this dimensionless
parameter, related to the fact that the effective 't~Hooft coupling
constant, $g_{YM}^2 K$, always becomes large in these theories at
high energies. Since the computations of short-distance correlation
functions that we will perform will be dominated by this asymptotic
region, our results will be valid (at short enough distances) for
any value of $g_s P$.

In the AdS/CFT correspondence \cite{m9711,AdS2point,Witten9802} (see
\cite{m2} for a review) properties of the conformal field theory
(CFT) may be computed using its holographic dual theory on
anti-de Sitter (AdS) space. In particular, many
computations can be done when the supergravity approximation is
valid. The same is true also for the cascading gauge theories
\cite{kras1,kras2,ABY,Berg:2005pd}. It was shown in \cite{ABY} that
despite having an infinite number of high-energy degrees of freedom,
all one-point functions of the cascading gauge theory (including the
conformal anomaly) are finite after they are holographically
renormalized. The consistency of this renormalization procedure was
tested in \cite{cashydro}. It can be used to compute the
thermodynamic properties, such as the pressure and the energy
density, of the cascading gauge theory plasma, by evaluating the
one-point function of the stress-energy tensor. From these one can
compute the speed of sound in the theory. The same speed of sound is
precisely reproduced from the dispersion relation for the sound
waves extracted from the pole in the stress-energy tensor two-point
correlation function, which can be evaluated without holographic
renormalization.

In this paper we continue exploring the holographic definition of
cascading gauge theories. We compute the large-momentum limit of a
specific contribution to the $N$-point functions of
operators dual to supergravity fields. We show that this
contribution has only analytic ultraviolet divergences (contact
terms) and, therefore, it
may be renormalized with local counterterms. We
demonstrate that holographic renormalizability of the $N$-point
correlation functions in the cascading theories is directly linked
to the renormalizability of the corresponding conformal field
theories (which arise in the $P\to 0$ limit). For example,
this implies that the
renormalizability of the Klebanov-Strassler cascading
theory \cite{KT,ks} follows
from the renormalizability of the Klebanov-Witten supersymmetric CFT
\cite{kw}. Moreover, we find that properly normalized cascading
gauge theory operators have the expected scaling of their
$N$-point correlation functions. They behave as $N_{eff}^{2-N}$ with
$N_{eff}\propto \ln(k/\Lambda)$, where $k$ is a typical momentum
scale, as expected in the 't Hooft large $N_{eff}$ limit \cite{thooft}
of an $SU(N_{eff})$ gauge theory (or an $SU(N_{eff})\times
SU(N_{eff}+P)$ gauge theory).
Our analysis confirms the
interpretation of the cascading gauge theories as renormalizable
four-dimensional quantum field theories whose effective number of
degrees of freedom logarithmically increases with the energy, as
suggested also by their thermodynamic behavior at high
temperatures \cite{bkt1,bkt2,bkt3}.

This paper is organized as follows. We begin in section 2 by
outlining the computation that we need to perform. In section 3 we
compute the bulk-to-boundary propagator in the cascading background,
and use it to compute the two-point functions of the cascading
theories (some specific cases of two-point functions were previously
computed in \cite{kras1,kras2}). The main computation of the
$N$-point correlation functions is performed in section 4, and our
main results appear in \S4.2.4. Two
appendices contain technical details.

Our analysis leaves some remaining open problems. First,
we were only able to compute the large-momentum limit of a specific
class of correlation functions. It would be interesting to find a way to
generalize our computation to arbitrary correlation functions.
It would also be interesting to extend our results
to finite values of the momentum. Such correlation functions depend
on the IR behavior of the cascading theory, which our
large-momentum computation
is independent of. In particular, the computation of correlators
at finite momentum is necessary in order to compute the S-matrix
of the cascading theories (through an LSZ-type procedure), and
without it we are not able to say how the S-matrix of these theories
behaves at high energies.

Second, we only
discuss here correlation functions of the cascading gauge theory
operators which are dual to ten dimensional supergravity
modes\footnote{This includes in particular the operators dual to the
specific Kaluza-Klein modes appearing in the effective 5d
supergravity description of the asymptotically cascading geometries,
obtained in \cite{ABY}.}. It would be very interesting to evaluate
also the correlation functions of the operators dual to massive
string modes, and to verify whether or not they confirm the picture
of the cascading theory as a renormalizable quantum field theory
with $N_{eff}\propto \ln(k/\Lambda)$ effective degrees of freedom.
Note that this computation is
complicated since the
anomalous dimension of the corresponding operators seems to grow as
$(g_{YM}^2 N_{eff})^{1/4}$, which grows without bound as the
energy is increased.
A first important step in this direction was made in \cite{martin}
where it was shown that the anomalous dimension of twist-$2$
operators in cascading gauge theories has the expected dependence on
$N_{eff}$.

\section{Generalities}
We would like to study the high momentum correlation functions of
cascading gauge theories. Specifically, we look at the cascading
gauge theory of fractional D3-branes at a conifold singularity,
whose gravitational dual is asymptotically given by the
Klebanov-Tseytlin (KT) \cite{KT} solution of type IIB
supergravity\footnote{There are also
some $p$-form fields turned on in the background, which will not
play any role in our computations.}
\begin{equation}
    {1\over L^2} ds_{10}^2=g_{\mu\nu}dy^{\mu}dy^{\nu} + \sqrt{h}
        ds_{T^{1,1}}^2
=\frac{1}{\sqrt{h}}\frac{1}{\rho^2}
        {dx^i}^2+\frac{\sqrt{h}}{\rho^2} d\rho^2 + \sqrt{h}
        ds_{T^{1,1}}^2,
\eqlabel{assKT}
\end{equation}
where $h(\rho) = \frac{1}{8} P^2 p_0 + \frac{1}{4} K_0 -\frac{1}{2}
P^2 p_0 \ln (\rho/\rho_0)$ in the notation of \cite{ABY}, $\mu,\nu=
0,1,2,3,4$, $i=0,1,2,3$, $p_0$ is the string coupling
and $P$ is the number of fractional
D3-branes. We note that the details of this background (such as the
$T^{1,1}$ metric) will not play any role in our computations. These
should be valid for any cascading gauge theory background which has
a similar logarithmic
form of the warp factor $h(\rho)$. In order to keep our
discussion general we will denote $h(\rho) = a' - b
\ln(\rho/\rho_0)$. The AdS background is then a special case of
this with $b=0$.

As usual, we will perform a Kaluza-Klein (KK) reduction of all the
fields on the compact space $T^{1,1}$. After this reduction we
obtain an infinite set of five dimensional fields $\phi_i(x,\rho)$,
which are dual to operators $\mathcal{O}_i(x)$ in the dual field
theory.

In the AdS/CFT correspondence \cite{m9711,AdS2point,Witten9802} and
its generalizations, one obtains the field theory correlation
functions
$\langle\mathcal{O}_1({x_1})\ldots\mathcal{O}_N(x_N)\rangle$ in the
gravity approximation by varying the gravity action with respect to
the boundary value of the fields $\phi_i$ dual to
$\mathcal{O}_i(x_i)$. One contribution to such a correlation
function will come from an interaction vertex in the bulk action (if
it exists) of the form $\lambda_N \phi_1 \ldots \phi_N$.

A simple way to compute the correlation functions is to define a
bulk-to-boundary (BtB) propagator for the fields $\phi_i$, denoted
$K_i(x,x^{\prime};\rho)$, which gives the change in the field
$\phi_i(x^{\prime},\rho)$ in response to a (appropriately
normalized) delta function source at the point $x$ on the boundary
($\rho \to 0$). In section 3 we will review the form of these BtB
propagators in AdS space, and compute them for the cascading
background.

If we have a vertex of the form $\lambda_N \phi_1 \cdots \phi_N$
in the five dimensional effective action, we obtain by
the usual Feynman diagram techniques a contribution to the correlation
function of the form
\begin{equation}
    \langle\mathcal{O}_1({x_1})\ldots\mathcal{O}_N(x_N)\rangle
    =
    \lambda_N \int \left(\prod_{i=1}^N K_i(x_i,x;\rho)\right)
    h^{5/4}(\rho)\ \sqrt{g}\ d^{4}x d\rho,
\end{equation}
where $g$ is the five dimensional metric appearing in \eqref{assKT} and
the factor $h^{5/4}$ comes from the determinant of the $T^{1,1}$
metric.  We have absorbed some normalization factors
(including
the overall scale $L$) into the variation of the action
with respect to the boundary values of the fields.

A five dimensional interaction vertex of this form would arise from an
interaction between $N$ scalar fields in ten dimensions (with
$\phi_i$ the KK modes of these scalar fields).
However,
there are no such interactions in the 10d supergravity (SUGRA)
action. All the interactions in the 10d action involve two
derivatives. Namely, in 10d they look like a product of two
derivatives of 10d fields times a product of other fields. When we
reduce to 5d, we can get non-derivative interactions of the type
described above from ten-dimensional interactions with the
derivatives in the angular ($T^{1,1}$) directions; however, we then
get an additional factor of $h^{-1/2}(\rho)$ in the interaction vertex,
coming from the metric in the angular coordinates.
Thus, the
non-derivative interactions in the 5d SUGRA action which actually
exist take the form $\lambda_N h^{-1/2}(\rho) \phi_1 \cdots \phi_N$,
and their contribution to correlation functions takes the form
\begin{equation}
    \langle\mathcal{O}_1({x_1})\ldots\mathcal{O}_N(x_N)\rangle
    =
    \lambda_N \int \left(\prod_{i=1}^N K_i(x_i,x;\rho)\right)
    h^{3/4}(\rho)\ \sqrt{g}\ d^{4}x d\rho.
\eqlabel{npoint}
\end{equation}

Next, using
\begin{equation}
\sqrt{g}=\r ^{-5} h^{-3/4} \eqlabel{det5E},
\end{equation}
and Fourier transforming both sides of \eqref{npoint}, we find that
the momentum space correlation functions may be written using the
momentum space BtB propagators $\hat{K}_i$
(using translational and rotational
symmetry, which implies that the BtB propagators only depend on the
absolute value of their momentum) in the form
\begin{equation}
    \langle\hat{\mathcal{O}}_1({\vec{k}_1})\cdots
\hat{\mathcal{O}}_N(\vec{k}_N)\rangle
    =
    \delta(\sum_{i=1}^N \vec{k}_i)\int \prod_{i=1}^N
\hat{K}_i(|\vec{k}_i|;\rho) \rho^{-5} d\rho, \eqlabel{EnpointAdS}
\end{equation}
where $\hat{\mathcal{O}}_i(\vec{k}_i)$ is the
Fourier transform of $\mathcal{O}_i(x_i)$.
The large $k$ behavior can then be extracted from
this. We concentrate in this paper on non-derivative interactions in
five dimensions. Derivative interactions in the five dimensional
effective action can be treated similarly (they give expressions
similar to \eqref{EnpointAdS} but with $\rho$-derivatives of some of
the propagators or with additional factors of $k_i$) and do not
result in any qualitatively new physics.

The expression \eqref{EnpointAdS} is IR divergent as
$\rho \to 0$ (this corresponds to a UV divergence in the
field theory) and must be
renormalized. To regulate the theory, we put a cutoff
at some small (close to the boundary) radial coordinate $\rho_{UV}$,
and define a regularized bulk-to-boundary propagator corresponding
to a source at $\rho_{UV}$ instead of at the boundary.
In addition, the integral over the $\rho$ coordinate in
\eqref{EnpointAdS} extends from $\rho_{UV}$ to infinity. Eventually,
we need to take the $\rho_{UV}\to 0$ limit. Generically, the
integral \eqref{EnpointAdS} diverges as $\rho_{UV}$ goes to zero.
However, we will show that all the divergent terms are analytic in
(some of) the momenta, so they are contact terms in position space.

In the usual AdS/CFT correspondence, one can consistently subtract
these divergences using holographic renormalization, by adding
appropriate counter-terms to the action.  The final result for
$N$-point functions is
given by the non-analytic terms (in $k$) in the above expressions.
These terms
are non-divergent as $\rho_{UV}\to 0$ (if we are careful to take
the $\rho_{UV} \to 0$ limit only at the very end of
the calculation).
We will use precisely the same regularization and renormalization
method in the cascading background. We will find that this procedure
leads to finite results for the $N$-point correlation functions,
which are given by the non-analytic terms in \eqref{EnpointAdS} in
the $\rho_{UV}\to 0$ limit.

\section{Bulk-to-boundary propagators and two-point functions}
In this section we compute the BtB propagator in the
Klebanov-Tseytlin background \cite{KT} which is needed for the
computation of correlation functions. We specialize to the case of a
scalar field in five dimensions, coming from a KK reduction of some
type IIB supergravity field on the $T^{1,1}$. We expect the
generalization of our results to fields of higher spin to be
straightforward. We will find it more convenient to work in momentum
space
rather than in position space. Since holographic correlation
functions are usually computed in position space, we start in \S3.1
by reviewing the computation of the momentum space BtB propagator
in AdS space. This turns out to be useful because of the similarity
between the KT and AdS backgrounds. In \S3.2 we compute the BtB
propagator in the KT background, and in \S3.3 we use this for the
computation of two-point functions.

\subsection{The AdS bulk-to-boundary propagator}

We would like to find the BtB propagator for a scalar field of mass
$m$ moving in the $AdS_5$ background. This background is given by
setting $h=1$ in (\ref{assKT}), in which case the scale $L$ becomes
the radius of curvature (note that both $x$ and $\rho$ have units of
length). We assume that this scalar arises as some KK mode on
$T^{1,1}$ with the mass coming from the Laplacian in the $T^{1,1}$
directions. Plugging a solution with momentum $k^i$ in the $x^i$
directions into the equation of motion $(\square + m^2) \phi = 0$,
we find the equation
\begin{equation}
\label{AdSeq}
    \rho^2 (\rho^{-2} \phi)^{\prime\prime} + \rho (\rho^{-2} \phi)^{\prime} -
    (4+m^2 L^2+k^2\rho^2)(\rho^{-2}\phi) = 0.
\end{equation}
Here primes denote derivatives with respect to $\rho$, and $k\equiv
|\vec{k}|$.

Equation \eqref{AdSeq} is invariant under rescaling $\rho \to \alpha
\rho$, $k \to k / \alpha$, so the non-trivial features of the
solution will be at values of $\rho$ of order $1/k$. Now, suppose
that we solve the equation of motion in a space which is only
asymptotically AdS, with significant differences from AdS occurring at
$\rho > \rho_0$ (where $\rho_0$ is the scale where IR effects become
important). We expect that the solution to (\ref{AdSeq}) will be a
good approximation to the solution we are interested in as long as
$k \gg 1/\rho_0$. The corrections to this solution
will be a power series in $1/k \rho_0$, so our results will be valid at
large enough momentum in any asymptotically AdS space.

To solve (\ref{AdSeq}), we define $\psi \equiv \rho^{-2} \phi$ and
switch to dimensionless variables $R = \rho/\rho_s$ and $Y =
k\rho_s$, where $\rho_s$ is some arbitrary scale which we introduce
for convenience (it will, of course, drop out of all physical
results). In these variables the equation of motion takes the form
\begin{equation}
\label{E:AdSEOM}
    R^2 \psi^{\prime\prime}(R) + R \psi^{\prime}(R) - (\nu^2 + Y^2
    R^2)\psi(R) = 0,
\end{equation}
where $\nu^2=4+m^2 L^2$ is related to the dimension $\Delta$ of the
dual operator through $\Delta = \nu+2$. We restrict to $\nu
> 0$ such that we are strictly above the Breitenlohner-Freedman
bound for scalar fields in AdS space.

One method to solve the equation \eqref{E:AdSEOM}
is by an expansion at small $R$. In
the small $R$ limit there are two asymptotic solutions to the
equation of motion, $\psi \sim R^{\pm \nu}$. One can then expand the
solution as a power series in $R$ and in $\ln(R)$ around $R \sim 0$.
We find it convenient to write the two linearly independent
solutions as
\begin{align}
\label{E:Besselexp1}
    \psi_{\nu}(R) &= R^{\nu} \sum_{n=0}^{\infty} \tilde{\kappa}_{\nu,n} (Y R)^n
\intertext{and}
 \label{E:Besselexp1b}
    \psi_{-\nu}(R) &= R^{-\nu} \sum_{n=0}^{\infty}
\tilde{\kappa}_{-\nu,n} (Y R)^n
\end{align}
for non-integer $\nu$. For integer $\nu$ the second solution,
equation (\ref{E:Besselexp1b}), is replaced by
\begin{equation}
\label{E:Besselexp1c}
    \psi_{-\nu}(R) = R^{-\nu} \sum_{n=0}^{2\nu-1} \tilde{\kappa}_{-\nu,n} (Y R)^n
    + (-1)^{\nu+1} \ln \left( R \right) \psi_{\nu}(R).
\end{equation}

Before we impose any boundary conditions, the general
solution to the equations of motion is of the form
\begin{equation}
\label{E:CandalphaAdS}
 \phi=C_{\nu} (R^2 \psi_{-\nu}(R) + \alpha_{\nu} R^2 \psi_{\nu}(R)),
\end{equation}
where $C_{\nu}$ and $\alpha_{\nu}$ are arbitrary constants. The BtB
propagator $K(Y,R)$ is defined to be the solution with boundary
conditions such that it is finite in the interior, and such that it
equals $\rho_{UV}^{2-\nu}$ at the UV boundary $\rho = \rho_{UV}$
(this is the Fourier transform of the position-space boundary
condition $K(x_0,x;\rho) \to
\rho_{UV}^{2-\nu} \delta(x-x_0)$). $C_{\nu}$ is
easily determined by the UV boundary condition:
$C_{\nu}=\rho_{UV}^{2-\nu}(R_{UV}^2 \psi_{-\nu}(R_{UV}) +
\alpha_{\nu} R_{UV}^2 \psi_{\nu}(R_{UV}))^{-1}$, with $R_{UV} \equiv
\rho_{UV}/\rho_s$. On the other hand, $\alpha_{\nu}$ is determined
by the boundary condition in the interior of AdS. To find it, one
needs some handle on the asymptotic behavior of the solutions
$\psi_{\pm \nu}(R)$ at large $R$, which is not evident from the
series expansions we wrote above which are useful only when $R Y \ll
1$.

Fortunately, equation (\ref{E:AdSEOM}) is a Bessel equation and has
been thoroughly studied. The solution which is smooth as $R\to
\infty$ is given by a modified Bessel function of the second kind,
\begin{equation}
    {\hat K}(Y,R) =
\rho_{UV}^{2-\nu}\frac{R^2 K_{\nu}(R Y)}{R_{UV}^2 K_{\nu}(R_{UV} Y)}.
\end{equation}
This leads to
\footnote{
Note that the overall power of $Y$ is obvious from
dimensional analysis, as we know that the scale
$\rho_s$ should not appear in the solution, but this does not determine
the coefficient.}
\begin{equation}
    \alpha_{\nu} = \begin{cases}
        -Y^{2\nu}  & \nu\,\text{non-integer,} \\
        (-1)^{\nu+1} Y^{2\nu} \ln\left(\frac{1}{2} Y\right)& \nu\,\text{integer}.
    \end{cases}
\end{equation}

For later convenience, we give here the coefficients of the
small $R$ expansion of the modified Bessel function. We will find it
useful to write $K_{\nu}$ with a somewhat non-standard
normalization. We write
\begin{equation}
\label{E:Bessel_exp_I}
    K_{\nu}(y) = y^{-\nu}\biggl(\sum_{n=0}^{\nu-1}\kappa_{2n,0} y^{2n}
                +\sum_{n=\nu}^{\infty}\sum_{m=0}^1\kappa_{2n,m} y^{2n}\ln^m (y)\biggr)
\end{equation}
for integer $\nu$, and
\begin{equation}
\label{E:Bessel_exp_nI}
    K_{\nu}(y) = y^{-\nu}\sum_{n=0}^{\infty}\kappa_{2n,0} y^{2n}
                +y^{\nu}\sum_{n=0}^{\infty}\kappa_{2\nu+2n,0} y^{2n}
\end{equation}
for non-integer $\nu$. We choose $\kappa_{0,0}=1$.
The other coefficients are given in the following table. Note that
these $\kappa$'s are different from the ${\tilde \kappa}$'s in
equations (\ref{E:Besselexp1}) and (\ref{E:Besselexp1b}).

\begin{table}[hbtp]
\caption{\label{T:Bessel}  Coefficients of the modified Bessel
function of the second
    kind.}
\begin{center}
\begin{tabular}{|p{1in}|p{1in}|p{9cm}|}\hline
  $\nu$ integer & $n < \nu$   & $\kappa_{2n,0}=\frac{(-1)^n\Gamma(\nu-n)}{2^{2n}\Gamma(n+1)\Gamma(\nu)}$
        \\ \cline{2-3}
                & $n \geq \nu     $  & $\kappa_{2n,1} =
                                    \frac{(-1)^{\nu+1}}{2^{2n-1}\Gamma(\nu)\Gamma(n-\nu+1)\Gamma(n+1)}$
        \\ \cline{3-3}
                &             & $\kappa_{2n,0} =
                                     \frac{(-1)^{\nu} (\psi(n-\nu+1)+\psi(n+1))}{2^{2n}\Gamma(n-\nu+1)
                                        \Gamma(n+1)\Gamma(\nu)} -\ln(2)\kappa_{2n,1}$
        \\ \hline
        \multicolumn{2}{|l|}{$\nu$ non-integer}
                              & $\kappa_{2n,0}=\frac{(-1)^n\Gamma(\nu-n)}{2^{2n}\Gamma(n+1)\Gamma(\nu)}$
        \\ \cline{3-3}
        \multicolumn{2}{|l|}{}
                              & $\kappa_{2\nu+2n,0} =
                                    -\frac{\Gamma(1-\nu)}{2^{2n+2\nu}\Gamma(n+1)\Gamma(n+\nu+1)}$
        \\ \hline
\end{tabular}
\end{center}
\end{table}

\subsection{The KT bulk-to-boundary propagator}
Next we wish to solve for the BtB propagator in the KT background
(\ref{assKT}), for a scalar field arising as a KK mode on $T^{1,1}$.
We will use
the same notation as in (\ref{E:AdSEOM}), where now we can choose
$\rho_s = 1/\Lambda$
to be a typical IR scale in the cascading geometry.
The equation of motion that we find,
again writing the solution
as ${\hat K}(R) = R^2 \psi(R)$, is
\begin{equation}
\label{E:KTKKEOM}
    R^2 \psi^{\prime\prime} + R \psi^{\prime} -
    (\nu^2+Y^2 R^2\,h(R))\psi = 0,
\end{equation}
where $h(R)\equiv a'-b\ln(R\rho_s/\rho_0) = a-b\ln(R)$ (with $a
\equiv a' - b \ln(\rho_s/\rho_0)$).
It will sometimes be convenient to choose the scale $\rho_s$ such
that $a=0$.
Note that for KK modes, whose mass comes from
the Laplacian of some field on $T^{1,1}$, the factors of $h$ in
(\ref{assKT}) conspire such that $\nu$ appears in the equation in
exactly the same way as in AdS space.
Our analysis is an extension of the specialized analysis of the
$m^2=0$ case done in \cite{kras1,kras2}.

Of course, since the background (\ref{assKT}) is singular it is not
really meaningful to solve the equation of motion in it. We are
really interested in the solutions to the equations of motion in a
regular space which asymptotes to (\ref{assKT}), such as the
backgrounds found in \cite{ks} or in \cite{bkt1,bkt2,bkt3}. As in
the AdS case, we expect (and we can verify this based on our
results) that at large momentum $k$, the dominant contributions to
correlation functions will come from small values of $\rho$ of order
$1/k$ and that they will be independent of the IR (large $\rho$)
resolution of the
KT background. Thus, we will be interested in computing the large
$k$ limit of the BtB propagator and of the correlation functions,
which is universal to all asymptotically KT backgrounds. The details
of the IR resolution at a scale $\rho_0$
will affect corrections to the results of order
$1/k\rho_0$.

As in the AdS case, we start by finding a small $R$ expansion for
the solution to (\ref{E:KTKKEOM}), which we will denote by
$K^{(I)}$. For non-integer values of $\nu$, we write the solution as
a sum $\psi(R) \sim \psi_{\nu}(R) + \psi_{-\nu}(R)$ of two linearly
independent series expansions,
\begin{equation}
\eqlabel{EKTexpansion1}
    \psi_{-\nu}(R) = R^{-\nu} \sum_{\substack{n=0 \\ m \le
    n}}^\infty p_{2n,m} R^{2n} \ln^m (R),\qquad
\psi_{\nu}(R) = R^{\nu} \sum_{\substack{n=0 \\ m \le
    n}}^\infty p_{2n+2\nu,m} R^{2n} \ln^m (R).
\end{equation}
For integer values of $\nu$ it is simpler to write the two independent
solutions in a single series
\begin{equation}
    \psi(R) = R^{-\nu} \biggl\{\sum_{n=0}^{\nu-1}\sum_{m=0}^n\  p_{2n,m} R^{2n} \ln^m (R)+
\sum_{n=\nu}^{\infty}\sum_{m=0}^{n+1}\  p_{2n,m} R^{2n} \ln^m (R)
\biggr\}. \eqlabel{EKTexpansion3}
\end{equation}
Plugging these ansatze into the equation of motion
(\ref{E:KTKKEOM}), one obtains a recursive relation for the
coefficients $p_{2n,m}$ which is written and solved in appendix
\ref{A:AppKTKKsolution}. One finds that, both for integer and for
non-integer $\nu$, two of the coefficients are undetermined. One can
choose these to be $p_{0,0}$ and $p_{2\nu,0}$. This behavior is
analogous to the AdS case (see equations (\ref{E:Besselexp1}) -
(\ref{E:Besselexp1c})), where we denoted the undetermined constants
by $C_{\nu}$ and $\alpha_{\nu}$. The coefficients of the leading
power of $\ln(R)$, appearing at each order in the expansion of the
solution in powers of $R$, are summarized in table \ref{T:KT}.

\begin{table}[hbtp]
\caption{\label{T:KT} Leading coefficients in the expansions
    \eqref{EKTexpansion1} and \eqref{EKTexpansion3}. See \eqref{int1}-\eqref{int4}
for integer $\nu$, and \eqref{negnu}-\eqref{negnu3} for non-integer
$\nu$. The expression for the integer $\nu$, $n=\nu+s$ case includes
only the dependence of $p_{2\nu+2s,s}$ on $p_{2\nu,0}$ and does not
include its dependence on $p_{0,0}$.}
\begin{center}
\renewcommand{\arraystretch}{1.25}
\begin{tabular}{|p{1in}|p{1in}|p{8cm}|}\hline
  $\nu$ integer & $1\le n <    \nu$   & $p_{2n,n}=\frac{(bY^2)^{n}\Gamma(\nu-n)}{2^{2n}\Gamma(n+1)\Gamma(\nu)}\ p_{0,0}$
                                  \\
        \cline{2-3}
                & $n=\nu     $   & $p_{2\nu,\nu+1} =
                                    -\frac{(bY^2)^{\nu}}{2^{2\nu-1}\Gamma(\nu)\Gamma(\nu+2)}\ p_{0,0}$
                                \\
        \cline{2-3}
                & $n > \nu$ & $p_{2n,n+1}=\frac{(-1)^{n+\nu+1}
                    (bY^2)^n}{2^{2n-1}(\nu+1)\Gamma(\nu)\Gamma(n-\nu+1)\Gamma(n+1)}\
                    p_{0,0}$
                                 \\
        \cline{2-3}
                & $n=\nu+s $ & $p_{2\nu+2s,s} \cong
                    \frac{(-bY^2)^s\Gamma(\nu+1)}{2^{2s}
                    \Gamma(s+1)\Gamma(s+\nu+1)} p_{2\nu,0}$ \\
        \hline
  $\nu$ non-integer &  $n\ge 1$   & $p_{2n,n}=\frac{(bY^2)^{n}\Gamma(\nu-n)}{2^{2n}\Gamma(n+1)\Gamma(\nu)}\ p_{0,0}$
                                  \\
        \cline{2-3}
                    &  $n\ge 1$   & $p_{2\nu+2n,n} =
                                     \frac{(-b Y^2)^n \Gamma (1+\nu)}{2^{2n} \Gamma(n+1)\Gamma(n+1+\nu)}\ p_{2\nu,0}$
                                 \\
        \hline
\end{tabular}
\end{center}
\end{table}

To find the BtB propagator we need to determine the two integration
constants.
One of the constants is determined
from the UV boundary condition which we choose to be the same as in
AdS,
\begin{equation}
{\hat K}^{(I)}(R_{UV}) =
R^2 \psi(R)\bigg|_{R=R_{UV}}\ =\ \r^{2-\nu}_{UV},
\eqlabel{overnorm}
\end{equation}
while the other one is fixed by requiring
that the propagator is non-singular everywhere in the interior.
To find the latter coefficient, we need some handle on
the asymptotic behavior of the solutions as $\rho$ becomes very
large. This cannot be obtained from the perturbative expansion we
have given here, as this expansion is valid only in the region
(which we call region I) where
\begin{equation}
\eqlabel{region1}
    Y^2 R^2 \ln (R) \ll 1.
\end{equation}
In order to find the correct
integration constants
we will use the method of Krasnitz \cite{kras1,kras2}, which is to
solve the equations of motion in a region which allows for an
evaluation of the asymptotic value of the field and also has some
overlap with region I \eqref{region1}.

Consider the following approximation:
\begin{equation}
    h(R) = a - b \ln\left(\frac{R Y}{Y}\right)
        =a+b\ln(Y)\left(1-\frac{\ln(R Y)}{\ln(Y)}\right)
        =h_Y\left(1-\frac{b\ln(R Y)}{h_Y}\right),
\eqlabel{hydef}
\end{equation}
where we defined $h_Y \equiv h\left(\ft 1Y\right) = a + b \ln(Y)$.
We would like to solve the equation of motion (\ref{E:KTKKEOM}) in a
region where one has $h(R) \simeq h_Y$. We require
\begin{equation}
    |b\ln (RY)| \ll h_Y,
\end{equation}
which for large momentum, $Y\gg 1$, means
\begin{equation}
|\ln (RY)|\ll \ln (Y). \eqlabel{region2}
\end{equation}
In this region (which we call region II) we may
approximate the equation of motion (\ref{E:KTKKEOM}) as
\begin{equation}
    R^2 \psi^{\prime\prime} +
    R\psi^{\prime}-(\nu^2+R^2 Y^2 h_Y)\psi=0
\eqlabel{eqreg2}
\end{equation}
which is simply a Bessel equation. This Bessel equation has two
independent solutions. One solution is finite when its argument is
large and the other solution diverges. Thus, if this equation was
valid at large $R$ we would have chosen the solution
\begin{equation}
    {\hat K}^{(II)} = B R^2\ K_{\nu}(R Y \sqrt{h_Y})
\eqlabel{Kii}
\end{equation}
with some undetermined constant $B$.
Of course, at large $R$ we do not really trust this equation since
we are no longer in the region (\ref{region2}). However, as
discussed above, we expect that at large momentum the dominant
contributions to the correlation functions will come from regions
with $R Y$ which is not very large, and it should not matter what
the solution (or the background) looks like at large $R$. So, we
will choose the specific solution (\ref{Kii}) in region II, assuming
that even if we also include the other solution with some
coefficient (which will be present for generic IR resolutions of the
background) it will not change the leading large momentum behavior.

Notice that for $Y\gg 1$ there is an overlap between  region I
\eqref{region1} and region II \eqref{region2}.
 Indeed, if for large $Y$ we look at values of $R$ scaling as
\begin{equation}
R\sim \frac{1}{Y \ln^{\gamma} (Y)},\qquad \gamma>\frac 12,
\eqlabel{overlap}
\end{equation}
we are simultaneously in both regions. We would like
to exploit this overlap between regions I  and II
 to determine the coefficients $\{p_{0,0},p_{2\nu,0},B\}$ by matching
$K^{(I)}$ and $K^{(II)}$ in the overlap region. We will treat
the cases of non-integer and integer $\nu$ separately.

\subsubsection{Non-integer $\nu$}

From the UV boundary condition \eqref{overnorm} we have
(defining a normalized integration constant $C_{\nu}$)
\begin{equation}
p_{0,0} \equiv \r_{s}^{2-\nu} C_\nu
= \r_{s}^{2-\nu}(1+\mathcal{O}(R_{UV}^2 \ln(R_{UV}))).
\eqlabel{p00ni}
\end{equation}
Next, comparing the coefficients of terms going as $R^{2n-\nu}$ (for
integer $n \geq 0$) in the expansion (\ref{EKTexpansion1}) in region
I with the expansion of the Bessel function in region II, in the
overlap region \eqref{overlap}, we find
\begin{equation}
\begin{split}
R^{2n-\nu}\ B \k_{2n,0} Y^{2n-\nu}h_Y^{n-\frac \nu2}
\simeq  &R^{2n-\nu}\ \sum_{m=0}^n\ p_{2n,m}\ln^m(R)
\\
=&R^{2n-\nu}\ \sum_{m=0}^n\ p_{2n,m}(\ln(RY)-\ln(Y))^m\\
=& R^{2n-\nu}\  \biggl(p_{2n,n} (-1)^n \ln^n (Y)+\calo(\ln(RY)\
\ln^{n-1}(Y))\biggr),
\end{split}
\eqlabel{compare}
\end{equation}
where in the bottom line we used \eqref{region2}. Since we can use
the scaling \eqref{overlap}, we find
\begin{equation}
\begin{split}
B=&\frac{p_{2n,n}}{\kappa_{2n,0}}\
(-1)^n Y^{\nu-2n} h_Y^{\frac \nu2-n} \ln^n (Y)\times
\left(1+\calo\left({\ln(\ln(Y))\over \ln(Y)}\right)\right)\\
=&p_{0,0}\ \left(Y\sqrt{h_Y}\right)^\nu\ \times
\left(1+\calo\left({\ln(\ln(Y))\over \ln(Y)}\right)\right).
\end{split}
\eqlabel{bni}
\end{equation}
Note that this result is independent of $n$ (which is a consistency
check for the validity of both expansions).
Similarly, comparing the coefficients of terms going as $R^{2n+\nu}$ gives
\begin{equation}
\begin{split}
R^{2n+\nu}\ B \k_{2\nu+2n,0} Y^{2n+\nu}h_Y^{n+\frac \nu2}\simeq &R^{2n+\nu}\
\sum_{m=0}^n\ p_{2n+2\nu,m}\ln^m(R)\\
=& R^{2n+\nu}\ \sum_{m=0}^n\ p_{2n+2\nu,m}(\ln(RY)-\ln(Y))^m\\
=& R^{2n+\nu}\  \biggl(p_{2n+2\nu,n} (-1)^n \ln^n(Y)+\calo(\ln(RY)
\ln^{n-1}(Y))\biggr),
\end{split}
\eqlabel{comparep}
\end{equation}
leading to
\begin{equation}
\begin{split}
B=&\frac{p_{2n+2\nu,n}}{\k_{2\nu+2n,0}}\ (-1)^n Y^{-\nu-2n}
h_Y^{-\frac \nu2-n} \ln^n(Y)\times
\left(1+\calo\left({\ln(\ln(Y))\over \ln(Y)}\right)\right)\\
=&-p_{2\nu,0}\ \frac{2^{2\nu}\Gamma(1+\nu)}{\Gamma(1-\nu)}\
\left(Y\sqrt{h_Y}\right)^{-\nu}\ \times
\left(1+\calo\left({\ln(\ln(Y))\over \ln(Y)}\right)\right)
\end{split}
\eqlabel{bnip}
\end{equation}
(which, again,
is independent of $n$). Given \eqref{bni} this determines
\begin{equation}
p_{2\nu,0}=-p_{0,0}\ \frac{\Gamma(1-\nu)}{2^{2\nu}\Gamma(1+\nu)}\
\left(Y\sqrt{h_Y}\right)^{2\nu}\ \times
\left(1+\calo\left({\ln(\ln(Y))\over \ln(Y)}\right)\right).
\end{equation}

To summarize, for non-integer $\nu$, matching $K^{(I)}$ and
$K^{(II)}$ in the overlap region determines (to leading order at
large $Y$, with corrections of order $\ln(\ln(Y))/\ln(Y)$)
\begin{equation}
\begin{split}
p_{0,0}=&\r_{s}^{2-\nu} C_{\nu}, \qquad C_{\nu} = 1 + \calo(R_{UV}^2
\ln(R_{UV})),\\
p_{2\nu,0}=&-\r_{s}^{2-\nu}\ \frac{\Gamma(1-\nu)}{2^{2\nu}\Gamma(1+\nu)}\
    \left(Y\sqrt{h_Y}\right)^{2\nu} C_{\nu},\\
B=&\r_{s}^{2-\nu}\ \left(Y\sqrt{h_Y}\right)^\nu C_{\nu}.
\end{split}
\eqlabel{sumnonint}
\end{equation}

\subsubsection{Integer $\nu$}

Now we tackle the slightly more difficult case of integer $\nu$.
Again, from \eqref{overnorm} we
have
\begin{equation}
p_{0,0}=\r_{s}^{2-\nu} C_{\nu},\qquad C_{\nu} = 1 + \calo(R_{UV}^2
\ln(R_{UV})). \eqlabel{p00i}
\end{equation}
The comparison in the overlap region of terms of order $R^{2n-\nu}$ for
$\nu>n\ge 0$ is exactly the same as before. This leads to
\begin{equation}
B= p_{0,0}\ \left(Y\sqrt{h_Y}\right)^\nu\ \times
\left(1+\calo\left({\ln(\ln(Y))\over \ln(Y)}\right)\right). \eqlabel{bi}
\end{equation}
For $n=\nu$ we find
\begin{equation}
R^{\nu}\ B\  Y^{\nu}h_Y^{\frac \nu2}\sum_{m=0}^1 \k_{2\nu,m}
\ln^m \left(RY\sqrt{h_Y}\right)
\simeq R^{\nu}\ \sum_{m=0}^{\nu+1}\ p_{2\nu,m}\ln^m(R)
\end{equation}
or
\begin{equation}
R^{\nu}\ B\  Y^{\nu}h_Y^{\frac \nu2}\biggl(\k_{2\nu,0}+\frac 12
\k_{2\nu,1}\ln(h_Y)+\k_{2\nu,1} \ln \left(RY\right)\biggr) \simeq
R^{\nu}\
\sum_{m=0}^{\nu+1}\ p_{2\nu,m}(\ln(RY)-\ln(Y))^m.
\eqlabel{comparepi}
\end{equation}
The coefficients $p_{2\nu,m}$ with $m > 0$ scale as $Y^{2\nu}
p_{0,0}$. Therefore, in the large $Y$ limit, contributions with $0 <
m < \nu+1$ are all smaller than the contribution coming from
$p_{2\nu,\nu+1}$. On the other hand, $p_{2\nu,0}$ is an independent
coefficient so that apriori we do not know if it is smaller.
Thus, to leading order in $\ln(RY)/\ln(Y)$,
the right-hand side of \eqref{comparepi} takes the form
\begin{equation}
R^\nu \biggl(p_{2\nu,0}+ (-1)^{\nu+1}p_{2\nu,\nu+1}\ln^{\nu+1}(Y)
\biggr) \times \left(1+\calo\left({\ln(RY)\over \ln(Y)}\right)\right).
\eqlabel{lhsin}
\end{equation}
Comparing the leading order expansion of the left-hand side of
\eqref{comparepi} with \eqref{lhsin}, we find
\begin{equation}
B\ Y^\nu h_{Y}^{\ft \nu2}\ \k_{2\nu,0}=
p_{2\nu,0}+(-1)^{\nu+1}p_{2\nu,\nu+1}\ln^{\nu+1}(Y),
\eqlabel{calint2}
\end{equation}
where we have neglected terms of order $\ln(\ln (Y))/\ln(Y)$. This
implies (using our result \eqref{bi} for $B$) that to leading order
in  $1/\ln(Y)$
\begin{equation}
p_{2\nu,0}= (-1)^{\nu}p_{2\nu,\nu+1}\ln^{\nu+1}(Y),
\end{equation}
with $p_{2\nu,\nu+1}$ given in terms of $p_{0,0}$ in table \ref{T:KT}.

At this stage all the free parameters $p_{0,0},p_{2\nu,0}$ and $B$
are determined.  Thus, for each value of $n>\nu$,
matching the leading coefficients at order $R^{2n-\nu}$ in the overlap
region must be automatic. We explicitly verified that this is indeed
the case.

To summarize, for integer $\nu$ a matching of $K^{(I)}$ and
$K^{(II)}$ in the overlap region determines (to leading order in
$\ln(\ln(Y))/\ln(Y)$)
\begin{equation}
\begin{split}
p_{0,0}=&\r_{s}^{2-\nu} C_{\nu},\qquad C_{\nu} = 1 + \calo(R_{UV}^2
\ln(R_{UV})),\\
p_{2\nu,0}=&
(-1)^{\nu}p_{2\nu,\nu+1}\ln^{\nu+1}(Y) C_{\nu},\\
B=&\r_{s}^{2-\nu}\ \left(Y\sqrt{h_Y}\right)^\nu C_{\nu}.
\end{split}
\eqlabel{sumint}
\end{equation}

\subsubsection{Explicit examples}

We give here some explicit examples of KT BtB
propagators (with $a=0$).

For the massless $\nu=2$ case our result is identical to the
result of Krasnitz
\cite{kras1,kras2} :
\begin{multline}
\label{BtBnu2}
    {\hat K}(Y,R)= C_2
    \biggl[1 + \frac{1}{4} b Y^2 R^2 \ln(R) + b^2 Y^4 R^4
    \biggl(-\frac{1}{128} \ln(R) + \frac{1}{64}\ln^2(R) -
    \frac{1}{48}\ln^3(R)
        \\- \frac{1}{48}\ln^3Y  +
    \mathcal{O}(\ln^2(Y)\ln(\ln (Y)))\biggr)
    +\mathcal{O}(R^6) \biggr].
\end{multline}

For $\nu=5/2$ we find
\begin{multline}
\label{E:BtBnu52}
    {\hat K}(Y,R)=C_{5/2} \rho^{-1/2}\biggl[1+
    b Y^2 R^2 \left( -\frac{1}{36} + \frac{1}{6} \ln(R)\right)
    +b^2 Y^4 R^4 \biggl( \frac{1}{16} + \frac{1}{18} \ln(R)
   \\ \qquad \qquad \qquad +\frac{1}{24} \ln^2(R)\biggr)
    -\frac{1}{45} Y^5 R^5 \left(h_Y^{5/2}+
\mathcal{O}(\ln^2(Y)\ln(\ln (Y)))\right)
        +\mathcal{O}(R^6)
    \biggr].
\end{multline}

For the tachyonic $\nu=1$ case we find
\begin{equation}
\begin{split}
    {\hat K}(Y,R)=&C_1 \rho \biggl[1+
    b Y^2 R^2 \left(\frac{1}{4}\ln(R) - \frac{1}{4} \ln^2(R)
    +\frac{1}{4} \ln^2 (Y) + \mathcal{O}(\ln(Y)\ln(\ln (Y)))\right)\\
    &\qquad \qquad +\mathcal{O}(R^4)
    \biggr].
\end{split}
\label{E:BtBnu1}
\end{equation}

\subsection{Two-point functions}

In any holographic background the two-point function may be
extracted from the UV behavior of the BtB propagator. Two-point
functions in AdS were studied in \cite{AdS2point,Witten9802}, and
two-point functions in KT were studied (for $m^2=0$) in
\cite{kras1,kras2}.  One subtlety that was emphasized in
\cite{Freedman} is that in order to get correct Ward identities one
should use a prescription for evaluating the correlator in which the
UV cutoff $\rho_{UV}$ is taken to zero only at the very end of the
calculation. This is the prescription we will follow. Alternatively,
one may use holographic renormalization \cite{hol1,hol2,hol3,hol4,
hol5,hol6,hol7,hol8,hol9,hol10,ABY} to calculate the two-point
functions. Adding local counterterms does not change the result.

For completeness, we will first describe how to obtain the two-point
functions in AdS and then move on to the KT case. The reader may
refer to \cite{Freedman,AdS2point} for details. Consider a scalar
field in Euclidean AdS space (with a cutoff at $\rho=\rho_{UV}$)
with the action
\begin{equation}
S = {1\over 2} \int d^{d+1}x \sqrt{g} (g^{\mu \nu} \del_{\mu} \phi
\del_{\nu} \phi + m^2 \phi^2).
\end{equation}
Evaluating the action (in momentum space) on a solution $\hat{\phi}$
to the equations of motion gives
\begin{equation}
    S = \lim_{\rho \to \rho_{UV}}
        \frac{1}{2}
        \int \frac{1}{\rho^3}
        \delta(\vec{k}+\vec{q})\hat{\phi}(\vec{k})
        \partial_{\rho} \hat{\phi}(\vec{q})
        d^4k d^4q.
\end{equation}
The two-point function of the operator dual to $\phi$ is given by
the second derivative of the action with respect to sources
\begin{equation}
    \langle {\hat {\cal O}}_{\nu}(\vec{k})
{\hat {\cal O}}_{\nu}(\vec{q})\rangle
    =
    \delta(\vec{k}+\vec{q})\frac{1}{\rho_{UV}^3}
    \rho_{UV}^{2-\nu}\lim_{\rho\to\rho_{UV}}\partial_{\rho}
\hat{K}_{\nu}(\vec{q},\rho).
\end{equation}

This expression may be readily evaluated. We start with integer
$\nu$. Here, we have (see (\ref{E:Bessel_exp_I}))
\begin{multline}
    \partial_{\rho} \hat{K}_{\nu}(\vec{k}) =
    \rho_s^{-1} \partial_R \hat{K}_{\nu}
=    \rho_{s}^{1-\nu} C_{\nu}
    \Bigg(\sum_{n=0}^{\nu-1}\kappa_{2n,0}Y^{2n}(2n-\nu+2)R^{2n-\nu+1}
    \\
\qquad  +\sum_{n=\nu}^{\infty} \sum_{m=0}^1
\kappa_{2n+2\nu,m}Y^{2n+2\nu}
    ((\nu+2n+2)R^{2n+\nu+1}(\ln (RY))^m
+m R^{2n+\nu+1})
    \Bigg).
\end{multline}
So, the correlation function is given by the $R_{UV}\to 0$ limit of
\begin{multline}
    \delta(\vec{k}+\vec{q})
        \rho_{s}^{-2\nu}
R_{UV}^{-(\nu+1)} \times
        \\
\lim_{R\to R_{UV}} \biggl\{
        \left({\sum_{n=0}^{\nu-1}\kappa_{2n,0}(R_{UV} Y)^{2n}
           +\sum_{n=0}^{\infty}\sum_{m=0}^1\kappa_{2n+2\nu,m}(R_{UV} Y)^{2n+2\nu}(\ln (R_{UV} Y))^m
        }\right)^{-1}
        \\
\times
        \biggl(\sum_{n=0}^{\nu-1}\kappa_{2n,0}Y^{2n}(2n-\nu+2)R^{2n-\nu+1}
           \\
+\sum_{n=0}^{\infty}\sum_{m=0}^1\kappa_{2n+2\nu,m}Y^{2n+2\nu}
        \left((\nu+2n+2)R^{2n+\nu+1}(\ln (R Y))^m+m R^{2n+\nu+1}\right)
        \biggr)\biggr\}
\end{multline}
where the denominator comes from the normalization $C_{\nu}$. This
expression diverges as $\rho_{UV}\to 0$, but it is easy to verify
that all divergent terms are analytic in $k$. Analyticity of these
divergences implies that they are unphysical contact terms in
position space which may be subtracted (by adding appropriate
counter-terms). The finite non-analytic terms are given by
\begin{equation}
    \langle {\hat {\cal O}}_{\nu}(\vec{k})
{\hat {\cal O}}_{\nu}(\vec{q}) \rangle=
    \delta(\vec{k}+\vec{q}) \frac{(-1)^{\nu+1}}{2^{2\nu-2}\Gamma(\nu)^2}
        k^{2\nu} \ln (k\rho_s).
\end{equation}
Despite appearances, this is independent of $\rho_s$, since the
$\rho_s$-dependent term is analytic. Note that,
as stated earlier, if we had first taken the $\rho_{UV} \to 0$
limit, the numerical coefficient we would have obtained would have
been different (and wrong).

The analysis for non-integer $\nu$ is very similar, the only
difference is that instead of (\ref{E:Bessel_exp_I}) we have the
expansion (\ref{E:Bessel_exp_nI}). Since non-analyticity only comes from
the second sum, the results are similar,
\begin{equation}
    \langle {\hat {\cal O}}_{\nu}(\vec{k})
{\hat {\cal O}}_{\nu}(\vec{q}) \rangle=
\delta(\vec{k}+\vec{q}) k^{2\nu}(2\nu)\kappa_{2\nu,0}\
    =\delta(\vec{k}+\vec{q}) \frac{-\Gamma(1-\nu)}{2^{2\nu-1}\Gamma(\nu)}
        k^{2\nu}.
\end{equation}

To obtain the position space correlation function, we need to
Fourier transform the above expressions. We find that for both integer
and non-integer $\nu$ we have \cite{AdS2point,Freedman}
\begin{equation}
    \langle {\cal O}_\nu(\vec{x}_1) {\cal O}_{\nu}(\vec{x}_2)\rangle
    =
    \frac{2 \nu^2(1+\nu)}{\pi^2}\frac{1}{|\vec{x}_1-\vec{x}_2|^{4+2\nu}}.
\end{equation}

The analysis of the two-point functions in the cascading (KT)
background closely follows the AdS case and generalizes the results
of \cite{kras1} to massive fields. We still have
\begin{equation}
    \langle {\hat {\cal O}}_{\nu}(\vec{k})
{\hat {\cal O}}_{\nu}(\vec{q}) \rangle
    =
    \delta(\vec{k}+\vec{q})
    \lim_{\rho\to\rho_{UV}}
    \frac{1}{\rho^3}\hat{K}_{\nu}(\vec{k})
\partial_{\rho}\hat{K}_{\nu}(\vec{q}).
\end{equation}
The leading non-analytic
contribution to this expression is similar to that in the
AdS case because of the form of the power law expansion of the
propagator (see (\ref{EKTexpansion1}) and (\ref{EKTexpansion3})). We
find
\begin{equation}
    \langle {\hat {\cal O}}_{\nu}(\vec{k})
{\hat {\cal O}}_{\nu}(\vec{q}) \rangle
    =
    \delta(\vec{k}+\vec{q})\rho_s^{-2-\nu}(2\nu)p_{2\nu,0}.
\end{equation}
For integer $\nu$ this gives us
\begin{equation}
    \langle {\hat {\cal O}}_{\nu}(\vec{k})
{\hat {\cal O}}_{\nu}(\vec{q}) \rangle
    =
    \delta(\vec{k}+\vec{q})\frac{(-1)^{\nu+1} b^{\nu}}{2^{2\nu-2}
(\nu+1)\Gamma(\nu)^2}
    k^{2\nu}(\ln (Y))^{\nu+1},
\end{equation}
while for non-integer $\nu$ we find
\begin{equation}
    \langle {\hat {\cal O}}_{\nu}(\vec{k})
{\hat {\cal O}}_{\nu}(\vec{q}) \rangle
    =
    -\delta(\vec{k}+\vec{q})\frac{\Gamma(1-\nu)}{2^{2\nu-1}\Gamma(\nu)}
    k^{2\nu}(b \ln (Y))^{\nu}.
\end{equation}

By Fourier transforming we may obtain the leading behavior of the
short distance correlation function. We find
\begin{equation}
    \langle {\cal O}_\nu(\vec{x}_1) {\cal O}_{\nu}(\vec{x}_2)\rangle
    =
    \frac{2 \nu^2(1+\nu)}{\pi^2}\frac{\left(b
\ln(\rho_s/|\vec{x}_1-\vec{x}_2|)\right)^{\nu}}
    {|\vec{x}_1-\vec{x}_2|^{4+2\nu}}
    =
    \frac{2 (-b)^{\nu} \nu^2(1+\nu)}{\pi^2}\frac{\left(
\ln(\Lambda |\vec{x}_1-\vec{x}_2|)\right)^{\nu}}
    {|\vec{x}_1-\vec{x}_2|^{4+2\nu}}
\end{equation}
for both integer and non-integer $\nu$. Curiously, in the KT case we
find that the momentum space two-point functions are not smooth in
$\nu$, while in AdS the $\nu \to n$ limit, with $n \in \mathbb{Z}$,
commutes with the Fourier transform. In both cases the position
space answers are smooth in $\nu$. Similarly, the full momentum space
KT BtB propagator does not seem to have a smooth limit as $\nu$
approaches an integer, while its Fourier transform does (at least
for the leading terms which we computed).

On general grounds, we expect the correlation functions of the
cascading gauge theories to reflect the variation in the rank of the
gauge group with the momentum. Since in the large $K$ limit of an
$SU(K)$ gauge theory, there is a standard normalization of the
operators (which is the one coming from the dual string theory) in
which all correlation functions scale as $K^2$, we expect to have a
normalization in the KT case for which all correlation functions scale
as $N_{eff}^2 \sim b^2 \ln^2(k/\Lambda)$ (more precisely, since some
factors of $\ln(k)$ disappear when we go to position space for integer
$\nu$, we expect the position space answers to scale as $b^2
\ln^2(\Lambda |\vec{x}_i-\vec{x}_j|)$).
The two-point functions we found above
have this scaling for the massless $\nu=2$ case, but not for other
cases. However, we can always rescale our operators (by a
momentum-dependent factor) so that the 2-point functions will all
scale as $b^2
\ln^2(\Lambda
|\vec{x}_i-\vec{x}_j|)$ as expected. Therefore, we will define
normalized operators
${\tilde {\cal O}}_{\nu}(\vec{k}) \equiv
{\hat {\cal O}}_{\nu}(\vec{k}) / (b \ln(k/\Lambda))^{(\nu-2)/2}$
which obey the expected
scaling for their 2-point functions. In the following section we
will compute general correlation functions of these normalized
operators and see that they scale as $N_{eff}^2$. Another natural
scaling which is often used is to divide the previous operators
by $N_{eff}$, so that the two-point functions do not scale while
higher $N$-point functions scale as $N_{eff}^{2-N}$. The operators
obeying this scaling are
${\tilde {\cal O}}_{\nu}^{\prime}(\vec{k}) \equiv
{\hat {\cal O}}_{\nu}(\vec{k}) / (b \ln(k/\Lambda))^{\nu/2}$.

\section{Higher $N$-point functions}

In this section we compute the large-momentum behavior of $N$-point
functions in asymptotically cascading backgrounds. We will focus on
the specific contribution to $N$-point functions coming from a
single non-derivative vertex which couples
the $N$ fields. We expect that the qualitative features that we find
will be present also in other contributions to the $N$-point
functions. We begin by performing
our analysis in asymptotically AdS backgrounds, both because
as far as we know this computation
has not been performed before in momentum space, and because many
features of the AdS computation carry over in a straightforward
manner to the cascading case.

In general, the expression for the correlators (both in the AdS case
and in the cascading case) is quite complicated, involving an
integral of Bessel functions which we do not know how to compute
exactly. However, we will be able to prove that the results for the
non-analytic terms in the correlators are always finite (independent
of the UV cutoff), so that the theory is well-defined. In some
special cases we will be able to write down a closed-form expression
for the
leading large momentum behavior of the
correlators. Separating
our computation
into regions I and II as we did in the discussion of the KT BtB
propagator,
we will show that in some cases
(both in AdS and in KT), the region I contribution is dominant at
large momentum, and can be explicitly computed.

\subsection{Tree level $N$-point functions in AdS}

\subsubsection{General expression for the $N$-point
functions}\label{AdSNcomp}

We are interested in computing the contribution to an $N$-point
function of operators dual to scalar fields, coming from a tree-level
diagram involving a single interaction vertex (with coefficient
$\lambda_N$) coupling these scalar fields together. The general rules
of computation in AdS space \cite{AdS2point,Witten9802} imply that the
result in momentum space is given by
\begin{equation}
  \cala_N=  \langle\hat{\mathcal{O}}_1({\vec{k}_1})\ldots\hat{\mathcal{O}}_N(\vec{k}_N)\rangle
    =
    \delta\left(\sum \vec{k_i}\right)\ \lambda_N\int \prod_i \hat{K}_i(k_i;\rho) \rho^{-5} d\rho.
\eqlabel{npads}
\end{equation}
As discussed above, the same expression should be true at high
momentum even in spaces which are only asymptotically AdS.

In AdS space we have an explicit result, described above, for the
BtB propagator $\hat{K}$ at any value of $\rho$. However, since we
are planning to generalize our results to the KT background, it is
natural to separate the contributions to \eqref{npads} as coming
from region I and region II, where in region I we use the
perturbative expansion of the BtB propagator
\eqref{E:Besselexp1}-\eqref{E:Besselexp1c}, which is useful for $R Y
\ll 1$, and in region II we use the precise expression involving the
Bessel function. These regions are analogous to the two regions we
used in our KT computation. In principle, in AdS we could extend
region II all the way down to $R=R_{UV}$. However, it will be
instructive to choose a separation point $R_t$ obeying $R_t Y \ll 1$
such that the power series expression can be used at $R < R_t$ and
the Bessel function expression can be used at $R > R_t$. In the KT
case we will be forced to use such a procedure.

In region I of AdS  we have seen that the BtB propagator is given by
\begin{equation}
 \hat{K}_{\nu}^{(I)}=   \r_s^{2-\nu} C_{\nu}
    R^{-\nu+2} \left(\sum_{n=0}^{\nu-1} \kappa_{2n,0} (R Y)^{2n}
    +
    \sum_{n=0}^{\infty}\sum_{m=0}^1
\kappa_{2n+2\nu,m}\ (RY)^{2n+2\nu}\ \ln^m(RY)\right)
\eqlabel{propAdS}
\end{equation}
for integer $\nu$, and by
\begin{equation}
 \hat{K}_{\nu}^{(I)}=   \r_s^{2-\nu}
    C_{\nu}
    Y^{\nu} R^2\left((RY)^{-\nu}\sum_{n=0}^{\infty} \kappa_{2n,0} (R Y)^{2n}
    +
    (RY)^{\nu} \sum_{n=0}^{\infty}
    \kappa_{2\nu + 2n,0}
    (R Y)^{2n}\right)
    \eqlabel{propAdSI}
\end{equation}
for non-integer $\nu$. The coefficients $\kappa$ are given in table
\ref{T:Bessel}. In order to unify our expressions for integer and
non-integer values of $\nu$, we will write both cases as
\begin{equation}
\hat{K}_{\nu}^{(I)} = \r_s^{2-\nu} C_{\nu} R^{-\nu+2} \sum_{n,m,s}
\kappa_{2n+2\nu s,m} (R Y)^{2 n + 2 \nu s} \ln^m(RY).
\end{equation}
In the sum, $n$, $m$ and $s$ take the following values. $s\in
\{0,1\}$ distinguishes the first and second terms in \eqref{propAdS}
and \eqref{propAdSI}. If $\nu$ is an integer and $s=1$ then $m\in
\{0,1\}$. Otherwise, $m=0$.  Finally, $n$ goes from zero to infinity
except when $\nu$ is an integer and $s=0$, in which case it goes
from zero to $\nu-1$.

Ignoring for now the momentum conserving
$\dd$-function, the tree level $N$-point function may be written as
\begin{equation}
\begin{split}
\cala_N=& \lambda_N \r_s^{-4}\ \int_{R_{UV}}^{\infty} dR\ R^{-5}\
\prod_{i=1}^N \hat{K}_i(Y_i,R) =
\cala_N^{(I)}+\cala_N^{(II)}\\
=&  \lambda_N \r_s^{-4}\ \int_{R_{UV}}^{R_t} dR\ R^{-5}\
\prod_{i=1}^N \hat{K}_i^{(I)}(Y_i,R)+ \lambda_N\r_s^{-4}\
\int_{R_{t}}^{\infty} dR\ R^{-5}\ \prod_{i=1}^N
\hat{K}_i^{(II)}(Y_i,R).
\end{split}
\eqlabel{an}
\end{equation}
We do not know how to perform the integral over the Bessel functions
in region II.
However,
in region I we can perform the integral explicitly:
\begin{equation}
\begin{split}
\cala_N^{(I)}=&\r_s^{-4}\ \lambda_N \prod_{i=1}^N C_{\nu_i}
\sum_{\{n_i,m_i,s_i\}}\ \biggl(\prod_{i=1}^N\ \r_s^{2-\nu_i}
\ Y_i^{2n_i+2s_i\nu_i}\ \k_{2n_i+2s_i\nu_i,m_i} \biggr)\\
&\times\ \int_{R_{UV}}^{R_t}\ \frac{dR}{R}\ R^{\nb}\prod_{j=1}^N\ln^{m_j} (RY_j)
\end{split}
\eqlabel{ai}
\end{equation}
where the summation is over $\{n_i,m_i,s_i\}$ in the range described
above, and we define
\begin{equation}
\nb\equiv -4+2N-\sum_{i=1}^N\nu_i+2\sum_{i=1}^N\nu_is_i
+2\sum_{i=1}^N
    n_i. \eqlabel{nbdef}
\end{equation}

In order to evaluate the integral in (\ref{ai}), we arrange the
indices such that all the $m_i$'s for which $m_i \neq 0$ appear
first, at $i=1,\cdots,\mb$ where $\mb\equiv \sum_{i=1}^N m_i$. We
note that $\mb$ counts the number of $\ln(Y_i)$ contributions from
integer $\nu$ terms. For $\nb \neq 0$ we find that the integral
\eqref{ai} is given by
\begin{equation}
    \int \prod_{i=1}^{\mb} \ln (Y_i R) R^{\bar{n}}
    \frac{dR}{R} =
    R^{\nb}
    \sum_{k=0}^{\mb} \frac{(-1)^k}{\nb^{k+1}}
    \sum_{\substack{j_1,\ldots,j_k \\ j_1 \neq\ldots\neq
    j_k}}^{\mb}\left( \prod_{i\neq j_1,\ldots,j_k} \ln (Y_i
    R)\right),
\label{E:logsintegral}
\end{equation}
where the second sum on the right is given by one when $k=0$. Thus,
the contribution of these terms is given by:
\begin{multline}
\cala_N^{(I)}=\r_s^{-4}\ \lambda_N \prod_{i=1}^N C_{\nu_i}
\sum_{\{n_i,m_i,s_i\},{\bar n}\neq 0}\ \biggl(\prod_{i=1}^N\
\r_s^{2-\nu_i}
\ Y_i^{2n_i+2s_i\nu_i}\ \k_{2n_i+2s_i\nu_i,m_i} \biggr)\\
\times
    R^{\bar{n}}
    \sum_{k=0}^{\mb} \frac{(-1)^k}{\bar{n}^{k+1}}
    \sum_{\substack{j_1,\ldots,j_k \\ j_1 \neq\ldots\neq
    j_k}}^{\mb}\left( \prod_{i\neq j_1,\ldots,j_k} \ln (Y_i
    R)\right)\Bigg|^{R_t}_{R_{UV}}.
\label{E:Aonebarnnonzero}
\end{multline}
For the special $\bar{n}=0$ case, we find
\begin{equation}
\label{E:nbarzerointegral}
    \int \prod_{i=1}^{\mb} \ln (Y_i R)
    \frac{dR}{R} =
    \sum_{t=0}^{\mb} \frac{(\ln (\mu
    R))^{t+1}}{(t+1)!}\partial^{(t)}P(-\ln(\mu)),
\end{equation}
where we have defined $P(x)\equiv \prod_{i=1}^{\mb} (\ln (Y_i)+x)$,
$\partial^{(t)}$ is the $t$'th derivative, and $\mu$ is an arbitrary
integration constant which should be independent of the momenta
$Y_i$.
Therefore, the $\bar{n}=0$ terms
contribute
\begin{multline}
\cala_N^{(I)}=\r_s^{-4}\ \lambda_N \prod_{i=1}^N C_{\nu_i}
\sum_{\{n_i,m_i,s_i\},{\bar n}=0}\ \biggl(\prod_{i=1}^N\
\r_s^{2-\nu_i}
\ Y_i^{2n_i+2s_i\nu_i}\ \k_{2n_i+2s_i\nu_i,m_i} \biggr)\\
\times
    \sum_{t=0}^{\mb} \frac{(\ln (\mu
    R))^{t+1}}{(t+1)!}\partial^{(t)}P(-\ln(\mu))
    \Bigg|^{R_t}_{R_{UV}}.
\label{E:Aonebarnzero}
\end{multline}

\subsubsection{Locality of UV divergences and the $R_{UV} \to 0$ limit}

Some of the terms in the integrals we wrote over region I (the ones
with ${\bar n} \leq 0$) are divergent as $R_{UV}\to 0$. The
correlation functions on AdS that we have been computing (which can
be, for example, those of the Klebanov-Witten supersymmetric gauge
theory \cite{kw}) should be renormalizable. Therefore, the
divergences in \eqref{ai} arising from the $R_{UV}\to 0$ limit must
be non-analytic in at most $(N-2)$ different momenta\footnote{All
such divergences can then be subtracted by introducing local
counter-terms of the type needed to renormalize all $\{2,\cdots,
N-1\}$-point correlation functions.}. This means that the
divergences do not contribute to the correlation functions in
position space at generic separated points and that they can be
canceled by local counter-terms. In our expressions for the
correlation function in region I, non-analytic contributions in
$k_i$ (or in $Y_i$) appear only when the corresponding $s_i=1$ (such
contributions are non-analytic due to the non-integer powers of
$k_i^2$ in the non-integer $\nu$ case, and due to the $\ln(Y_i)$ in
the integer $\nu$ case). Thus, we should require that \eqref{ai}
converges at the lower limit of integration whenever at least
$(N-1)$ of the $s_i$ are equal to one. The most stringent condition
comes from the case when a single $s_r=0$; we require that the
corresponding value of $\nb$ must be positive (for any choice of
$n_i$)
\begin{equation}
-4+2N+\sum_{i=1}^N\nu_i-2\nu_r> 0,\qquad
\nu_r\in\{\nu_1,\cdots,\nu_N\}. \eqlabel{nbrdef}
\end{equation}
Introducing
\begin{equation}
\nu_{max}\equiv \max\{\nu_1,\cdots,\nu_N\},\qquad \nu_{tot}\equiv
\sum_{i=1}^N\nu_i,
\end{equation}
we conclude that for the theory to be renormalizable, $\lambda_N$
must vanish whenever
\begin{equation}
-4+2N+\nu_{tot}-2\nu_{max}\le 0. \eqlabel{vanl}
\end{equation}
We can rewrite this condition in terms of the dimensions
$\Delta_i=\nu_i+2$ of the dual gauge theory operators. In this
language we find that $\lambda_N$ must vanish if
\begin{equation}
\frac 12 \sum_{i=1}^N\Delta_i\le \max\{\Delta_1,\cdots \Delta_N\},
\qquad {\rm or\ if\ for\ some\ }j\qquad \eqlabel{vanlg}
\Delta_j\ge \sum_{i=1,i\ne j}^N\Delta_i.
\end{equation}
This is, indeed, a well-known condition for renormalizability also
from the position-space analysis of AdS correlators \cite{extremal}.
The case with an equality in
\eqref{vanlg} is called the extremal correlator case, and the bulk
couplings $\lambda_N$ must vanish
in this case as well.

The condition described above holds in all known AdS backgrounds. In
particular, it holds for the KK modes in the Klebanov-Witten
background. We will see in the next subsection
that precisely when this condition
holds, the correlators of the same operators in the cascading KT
gauge theory
are also finite.
In this sense the renormalizability of the KT $N$-point correlation
functions is linked to the renormalizability of the corresponding
correlators in the conformal Klebanov-Witten gauge theory.

We also note that the $C_{\nu_i}$'s will not contribute to $N$-point
functions with $N\geq 3$, which implies that one may take the
$\rho_{UV}\to 0$ limit before evaluating the correlator (as shown in
\cite{Freedman}). This follows from the fact that
$\left(\prod_{i=1}^N C_{\nu_i} \right) =
 1+\mathcal{O}(R_{UV}^2)$ so that a non-trivial
contribution from the $C_{\nu_i}$
may survive in the $R_{UV}\to 0$ limit only if
there is a non-analytic term going as a negative power of $R_{UV}$.
As we have just shown, such divergent terms do not exist. Therefore,
when extracting the non-analytic contributions to the $N$-point
functions with $N\geq 3$ we may set $\prod_i C_{\nu_i} =1$.

\subsubsection{Analysis of leading terms}\label{leadads}

Evaluating the contribution to the correlation function \eqref{an}
from region II is technically difficult.
However, we will show that there are some $N$-point correlators
which are dominated by the $\bar{n}=0$ terms in the series expansion
in region I. We will evaluate these terms explicitly
below.
For more general correlators the best expression we have is
\eqref{an}.

We are interested in computing the correlation functions in the
large momentum limit. For simplicity, we assume that all momenta
$k_i$ are of the same order. This allows us to introduce a typical
momentum scale $k_* = Y_* / \rho_s$ with
\begin{align}
Y_{\star}=\frac{1}{N}\sum_{i=1}^N Y_i \eqlabel{typmom}.
\end{align}
When we perform an expansion of our expressions at
large $Y_{\star}$, we can ignore
terms of order $\ln(Y_i/Y_j)$ or $\ln(Y_i/Y_*)$ compared to terms of
order $\ln(Y_*)$.

In the large momentum limit, with $Y_i \sim Y_j$, we choose the
separation, $R_t$, between regions I and II to be
\begin{equation}
R_t=\frac{1}{Y_*\ln^\gamma (Y_*)},\qquad \gamma> 0,
\eqlabel{typmomc}
\end{equation}
such that $R_t Y_i \ll 1$ (namely, $R_t$ is in both regions I and
II). We note that \eqref{typmom} and \eqref{typmomc} imply that
$\ln(Y_*)$ and $\ln(R_t)$ are non-analytic in all momenta. There is
some freedom in choosing $R_t$ and $Y_{\star}$. However, in the large
momentum expansion, the final expressions we find will depend on the
choice of $R_t$ only through subleading terms. The specific choice
above for $R_t$ is motivated by the fact that in some exact
computations (such as a four-point correlation function which we
will present below) it correctly gives some of the subleading terms
as well.

We would like to find correlation functions whose major contribution
to non-analytic terms at high momenta is from region I, where we can
evaluate the integrals explicitly. We
note that at leading order in $Y_*$, the region II contribution is
equal to
\begin{equation}
\cala^{II} = \rho_s^{2N-4-\nu_{tot}} \lambda_N \int_{R_t}^{\infty}
R^{2N-4} \left(Y_*\right)^{\nu_{tot}} \prod _{i=1}^N
K_{\nu_i}\left(R Y_*\right) \frac{dR}{R} \eqlabel{compII}
\end{equation}
(up to a constant depending on the ratios of the momenta, which we
assume to be finite in the large momentum limit). Consider the case
$\nu_{tot}>2(N-2)$ \footnote{If $\nu_{tot} \leq 2(N-2)$
 then it can be shown that region II will always dominate over region I,
 and so, we are not interested in this case.}.
 Since $R_t Y_*$ is small in the limit we
 are interested in, and the small $R$ behavior of the integrand goes
 as $R^{2N-4-\nu_{tot}-1}$,
 the above integral is divergent as the
lower bound goes to zero. Hence, we conclude that it is dominated by
the contribution from the lower bound, which is of order
\begin{equation}
    \cala^{II} = \rho_s^{2N-4-\nu_{tot}} \lambda_N  \int_{R_t}
    \frac{dR}{R}\ R^{2N-4-\nu_{tot}}\sim \rho_s^{2N-4-\nu_{tot}} \lambda_N
    R_t^{2N-4-\nu_{tot}}.
\eqlabel{resII}
\end{equation}
We would like to compare this expression with the non-analytic
contributions from region I, which we computed in \S\ref{AdSNcomp}.
We start with the $\nb>0$ contributions. Approximating $Y_i \sim
Y_{\star}$ in (\ref{E:Aonebarnnonzero}) and requiring that it
dominate over the region II contribution gives us the condition
\begin{equation}
    \frac{(\gamma \ln(\ln (Y_{\star})))^{\mb}}
        {(\ln (Y_{\star}))^{\gamma(2\sum \nu_i s_i+2\sum n_i)}}=
 \frac{(\gamma \ln(\ln (Y_{\star})))^{\mb}}
        {(\ln (Y_{\star}))^{\gamma(\nb+\nu_{tot}-2(N-2))}}
    \gg
    1.
\end{equation}
Obviously this does not hold in the large momentum limit for any
$\gamma>0$.

Therefore, only terms with $\nb=0$ in region I
(\ref{E:Aonebarnzero}) may dominate. We saw that the contributions
from the lower bound of the integral are always analytic and thus,
uninteresting.
However, in the particular case of $\nb=0$,
the contribution to
\eqref{E:Aonebarnzero} from the upper region of integration is
non-analytic for any values of $s_i$, since it always contains a
$\ln (R_t)\sim -\ln (Y_*)$ term.
Recall that we are interested in the leading non-analytic
contribution at large momentum.
For integer $\nu_i$, every non-vanishing value of $s_i$ produces a
power of $\ln (Y_i)$ (from the term with $m_i=1$), so we would like
to have as many non-zero values of the $s_i$ corresponding to
integer $\nu_i$'s as possible. From the analysis of the previous
subsubsection, we know that there can be at most $(N-2)$ $s_i$'s
which do not vanish (for ${\bar n}=0$).

We find that the condition for the contribution from the upper bound
of integration of region I to dominate over the region II result is
\begin{equation}
    Y_{\star}^{4-2N+\nu_{tot}}
\sum_{t=0}^{\mb}
\left(\ln(\frac{\mu}{Y_{\star}\ln^{\gamma}(Y_{\star})})
        \right)^{t+1}
        \partial^{(t)} \prod_{i=1}^{\mb} (\ln (Y_i) +
        x)\big|_{x=-\ln(\mu)}
        \gg
        \frac{1}{(Y_{\star}\ln^{\gamma}(Y_{\star}))^{2N-4-\nu_{tot}}},
\end{equation}
or
\begin{equation}
    \sum_{t=0}^{\mb}
\left(\ln(\frac{\mu}{Y_{\star}\ln^{\gamma}(Y_{\star})})
        \right)^{t+1}
        \partial^{(t)} \prod_{i=1}^{\mb} (\ln (Y_i) +
        x)\big|_{x=-\ln(\mu)}
        \gg
    (\ln (Y_{\star}))^{\gamma(4-2N+\nu_{tot})}.
\end{equation}
This is satisfied at large momenta provided that
\begin{equation}
    \mb + 1 > \gamma(4-2N+\nu_{tot}).
\label{case21}
\end{equation}
We can always choose a small enough $\gamma$ (which must also satisfy
\eqref{typmomc}, $\gamma>0$) so that this inequality is satisfied.

To summarize, in order for an $N$-point function to be dominated
for large momentum by
the region I integral, we need that two constraints be satisfied.
One is a constraint on $\gamma$ which will make the contribution of
the $\nb=0$ term in region I dominate over region II (\ref{case21}).
It may always be satisfied. The other constraint is that a $\nb=0$
term should exist;
there should exist a choice of $s_i$ and $n_i$ such that
\begin{equation}
\label{E:casebnzero}
    -4+2N+\nu_{tot} + 2(\sum s_i \nu_i-\nu_{tot}) + 2 n_{tot}=0,
\end{equation}
with $n_{tot} \equiv \sum_i n_i$,
recalling that we must also have
\begin{equation}
\label{E:casebnzero2}
    -4+2N+\nu_{tot} - 2\nu_{max} > 0.
\end{equation}
Such a choice does not exist for all correlation
functions that we want to compute
(for example for generic non-integer values of
$\nu_i$). However, in many cases such a choice does
exist, and for any $N$ one can find some large enough
$\nu_i$ such that this constraint is satisfied.

\subsubsection{Examples of correlation functions}

We may now evaluate explicitly the correlation functions which are
dominated by the $\nb=0$  term in region I. These are correlation
functions which satisfy (\ref{E:casebnzero2}), and which have
contributions which satisfy (\ref{E:casebnzero}). In this case we
find that, to leading order in the momenta,
\begin{multline}
\label{E:exactAdS}
\langle {\hat {\cal O}}_1(\vec{k}_1)\cdots
{\hat {\cal O}}_N(\vec{k}_N) \rangle = \delta(\sum
\vec{k}_i)
    \lambda_N \sum_{\{n_i,s_i \in S \}}
    \left( \prod_{i=1}^N \kappa_{2n_i+2s_i\nu_i,m_i}\right)
    \left(\prod_{i=1}^N k_i^{2n_i+2s_i\nu_i}\right)
    \\
    \times
    \sum_{t=0}^{\mb}\frac{(-\ln (k_{\star}/\Lambda))^{t+1}}{(t+1)!}
        \partial^{(t)} \prod_{\substack{i=1 \\ s_i=1}}^{\mb}
        (\ln(k_i/\Lambda)+x)\Big|_{x=0},
\end{multline}
where $S$ is the set of all $s_i$'s and $n_i$'s which satisfy
(\ref{E:casebnzero}), and we take $m_i=1$ whenever $\nu_i$ is
integer and $s_i=1$. The constants $\kappa$ appear in table
$\ref{T:Bessel}$. We have used our freedom of choosing $\mu = 1/
(\Lambda\rho_s)$ to rewrite the logarithms using an arbitrary mass
scale $\Lambda$. The choice of $\mu$ does not affect any
non-analytic terms in the results since there are no
$\mu$-dependent terms which are non-analytic. When all
the $\nu_i$ are non-integer\footnote{Note that the Klebanov-Witten
background, unlike the $AdS_5\times S^5$ background, has KK modes
with non-integer values of the $\nu_i$.} (and also in other cases with
all $m_i=0$), the second line is simply given by
$-\ln(k_*/\Lambda)$.

As an  example consider 3-point correlation functions. Here we can
have at most a single $s_i\ne 0$. These correlators will have a
leading term which we can compute if there exist integer $n_i$'s
such that
\begin{equation}
\label{E:dominant1}
    2\sum_{i=1}^3 n_i = \sum_{i=1}^3 \nu_i-2\nu_j-2
\end{equation}
when $s_j=1$ for some $j$, or
\begin{equation}\label{E:dominant2}
    2\sum_{i=1}^3 n_i = \sum_{i=1}^3 \nu_i-2
    \end{equation}
when all $s_i=0$. In both cases we must also have
\begin{equation}
    0< 2+\sum_{i=1}^3 \nu_i - 2\nu_{max}.
\end{equation}
Defining $m_j = 1$ if $\nu_j$ is an integer, and zero otherwise, we
find from the (\ref{E:dominant1}) terms
\begin{multline}
\langle {\hat {\cal O}}_1(\vec{k}_1)\ldots
{\hat {\cal O}}_3(\vec{k}_3) \rangle_{a} =
\delta(\vec{k}_1+\vec{k}_2+\vec{k}_3)
    \lambda_N
    \sum_{j=1}^3
    \sum_{\{n_i\} \in S_j}\\
    \times
    \left(\prod_{i=1}^3 k_i^{2n_i}\right)
    k_j^{2\nu_j} (-\ln (k_{\star}/\Lambda) )
    \left(\ln (k_j/\Lambda)-\frac{1}{2}(\ln
    (k_{\star}/\Lambda))\right)^{m_j}
    \\
    \left( \prod_{i=1}^3
        \frac{(-1)^{n_i}\Gamma(\nu_i-n_i)}{2^{2n_i}\Gamma(n_i+1)\Gamma(\nu_i)}\right)
    \left(\frac{(-1)^{\nu_j}
        2}{\Gamma(1-\nu_j)\Gamma(\nu_j)}\right)^{m_j}
    \left(\frac{-\Gamma(n_j-\nu_j+1)}{2^{2\nu_j}\Gamma(n_j+\nu_j+1)}\right),
\label{E:AdS3point}
\end{multline}
where $S_j$ are all the $n_i$'s which satisfy (\ref{E:dominant1}).
Note that for integer $\nu_j$ the expression (\ref{E:AdS3point})
contains a ratio of diverging Gamma functions. This should be
understood as the finite limit of the ratio when $\nu_j$ approaches
the corresponding integer.  From the (\ref{E:dominant2}) terms we
find a contribution of
\begin{equation}
\langle {\hat {\cal O}}_1(\vec{k}_1)\cdots
{\hat {\cal O}}_3(\vec{k}_3) \rangle_{b} =
\delta(\vec{k}_1+\vec{k}_2+\vec{k}_3)
    \lambda_N
    \sum_{\{n_i\} \in S}
    \left(\prod_{i=1}^3
        \frac{(-1)^{n_i}\Gamma(\nu_i-n_i)}{2^{2n_i}\Gamma(n_i+1)\Gamma(\nu_i)}
        k_i^{2n_i}
    \right)
    \left(-\ln (\frac{k_{\star}}{\Lambda})\right),
\label{E:AdS3point2}
\end{equation}
where here $S$ are all the combinations of $\{n_i\}$ which satisfy
(\ref{E:dominant2}). The correlation function is generally given by
\begin{equation}
    \langle {\hat {\cal O}}_1(\vec{k}_1)\cdots
{\hat {\cal O}}_3(\vec{k}_3) \rangle
    =
    \langle {\hat {\cal O}}_1(\vec{k}_1)\cdots
{\hat {\cal O}}_3(\vec{k}_3) \rangle_{a}
    +
    \langle {\hat {\cal O}}_1(\vec{k}_1)\cdots
{\hat {\cal O}}_3(\vec{k}_3) \rangle_{b}.
\eqlabel{sumab}
\end{equation}
The expression \eqref{sumab} for the 3-point correlation function
should be understood as the leading non-analytic contribution for
fixed values of $\nu_i$, in the limit $\frac{k_i}{\Lambda}\to
\infty$, $\frac{k_*}{\Lambda}\to \infty$. For specific
choices\footnote{This occurs in particular when \eqref{E:dominant1}
can be solved for integer values of $\nu_i$'s.} of $\nu_i$'s, the
first contribution in \eqref{sumab}, \ie\ \eqref{E:AdS3point},
dominates. In this case the contribution \eqref{E:AdS3point2} is
subdominant, and it is inconsistent to keep it along with
\eqref{E:AdS3point}. It is only
when \eqref{E:AdS3point} and \eqref{E:AdS3point2} are of the same
order (in the large momentum limit) that the 3-point correlation
function is given by a sum \eqref{sumab}.

As a specific example, the three point massless ($\nu_i=2$)
correlator in momentum space is dominated at large momentum by the
terms with $s_j=1$ ($j=1,2,3$),
\begin{equation}
\langle {\hat {\cal O}}_2(\vec{k}_1)\ldots {\hat {\cal
O}}_2(\vec{k}_3) \rangle = \delta(\vec{k}_1+\vec{k}_2+\vec{k}_3)
    \frac{\lambda_N}{16}
       \sum_{j=1}^3
    k_j^4
    \left(\ln (k_{\star}/\Lambda) \ln (k_j/\Lambda)-\frac{1}{2}(\ln
    (k_{\star}/\Lambda))^2\right).
   \end{equation}

As a test of our methods we can look at the four-point function of
operators with indices $\nu_1=\nu_2=5/2$ and $\nu_3=\nu_4=1/2$. In
the half-integer $\nu$ case, modified Bessel functions of the second
kind are exponents multiplied by polynomials, and so the exact
momentum-space correlation function \eqref{an} can be evaluated
explicitly. From this explicit computation one finds that the
leading non-contact terms are given (up to the overall delta
function) by
\begin{equation}
\label{E:fourAdS}
        \frac{1}{6}\left(
        k_1^2+k_2^2-3(k_3+k_4)^2
        \right)
        \ln\left((k_1+k_2+k_3+k_4)/\Lambda\right).
\end{equation}
Our equation (\ref{E:exactAdS}) gives
\begin{equation}
    \frac{1}{6}\left(
        k_1^2+k_2^2-3(k_3+k_4)^2
        \right)
        \ln(k_{\star}/\Lambda),
\end{equation}
so that the leading large momentum non-analytic terms are indeed
identical.

Finally, we would like to emphasize that the general result
\eqref{sumab} is obtained in the $\frac{k_i}{\Lambda}\to
\infty$, $\frac{k_*}{\Lambda}\to \infty$ limit with $\nu_j$ kept
fixed. However, one can not use \eqref{sumab} to compute the 3-point
correlation functions arising in the limit as some $\nu_i$ approach
integer values. In fact, in the limit  $\nu_i\to \hat{\nu_i}$ for
integer values of $\hat{\nu_i}$ satisfying (\ref{E:dominant1}), one
has
\begin{equation}
\lim_{\nu_i\to \hat{\nu}_i}\ \langle {\hat {\cal O}}_{\nu_1}(\vec{k}_1)\cdots
{\hat {\cal O}}_{\nu_3}(\vec{k}_3) \rangle\ \ne\ \langle
{\hat {\cal O}}_{{\hat{\nu}}_1}(\vec{k}_1)\cdots
{\hat {\cal O}}_{{\hat{\nu}}_3}(\vec{k}_3) \rangle. \eqlabel{3pointlim}
\end{equation}
The reason for the apparent discrepancy is that the limit of the
short distance behavior of the correlation functions does not
commute with the limit in which dimensions of some operators
approach integer values. At the technical level, if the limit
$\nu_i\to \hat{\nu}_i$ is taken before the $\frac{k_i}{\Lambda}\to
\infty$ limit, certain $\nb\ne 0$ terms in \eqref{ai} can dominate
and produce a leading non-analytic behavior. As we explicitly
demonstrate in appendix \ref{limit}, the contribution of these
$\nb\ne 0$ terms in the limit $\nu_i\to \hat{\nu}_i$ precisely
reproduces the dominant $\nb=0$ contribution of the
$\hat{\cal O}_{\hat{\nu}_i}(\vec{k}_i)$
correlators in the large $k_i$ limit. Thus,
the correlation functions do have a smooth $\nu_i \to integer$
limit, as expected from the fact that the position space propagators
in AdS have a smooth limit.
However, this smooth limit is not evident in our expressions above.

\subsection{Tree level $N$-point functions in asymptotically cascading geometries}

\subsubsection{General expression for the $N$-point functions}

We would like to repeat the same analysis in asymptotically
cascading geometries. The bulk-to-boundary propagator in region I is
given by
\begin{equation}
 \hat{K}_{\nu}^{(I)}=
R^{-\nu+2} \left(\sum_{n=0}^{\nu-1}\sum_{m=0}^n p_{2n,m}
 R^{2n}\ln^m(R)
    +
    \sum_{n=0}^{\infty}\sum_{m=0}^{n+\nu+1}
p_{2n+2\nu,m}\ R^{2n+2\nu}\ \ln^m(R)\right) \eqlabel{propkt}
\end{equation}
for integer $\nu$, and by
\begin{equation}
 \hat{K}_{\nu}^{(I)}=
R^{-\nu+2}\left(\sum_{n=0}^{\infty} \sum_{m=0}^n p_{2n,m} R^{2n}\ln^m
 (R)
    +
     \sum_{n=0}^{\infty}\sum_{m=0}^{n}
    p_{2\nu + 2n,m}
    R^{2n+2\nu}\ln^m (R)\right)
\end{equation}
for non-integer $\nu$. The relevant coefficients, $p$, are given in
table \ref{T:KT}. Again, we can write both cases as
\begin{equation}
 \hat{K}_{\nu}^{(I)}=   R^{-\nu+2} \sum_{n,m,s} p_{2n+2\nu s,m}
 R^{2n+2\nu s}\ln^m(R),
 \end{equation}
where the only difference from the AdS case is in the range of $m$
(and in the precise coefficients).

Up to the momentum conservation $\dd$-function,   the tree level
$N$-point function arising from an $N$-point vertex in the bulk is
given by
\begin{equation}
\begin{split}
\cala_N=&\lambda_N \r_s^{-4}\ \int_{R_{UV}}^{\infty} dR\ R^{-5}\
\prod_{i=1}^N \hat{K}_i(Y_i,R) = \cala_N^{(I)}+\cala_N^{(II)}
\\
=&  \lambda_N \r_s^{-4}\ \int_{R_{UV}}^{R_t} dR\ R^{-5}\
\prod_{i=1}^N \hat{K}_i^{(I)}(Y_i,R)+ \lambda_N\r_s^{-4}\
\int_{R_{t}}^{\infty} dR\ R^{-5}\ \prod_{i=1}^N \hat{K}_i^{(II)}(Y_i,R),
\end{split}
\eqlabel{ankt}
\end{equation}
where $R_t$ is chosen to be in the overlap between regions I and II,
described in \eqref{overlap}. We are taking the background to be the
KT background all the way to $R\to \infty$ even though this is
singular, since we expect (and will verify) that the leading
contributions at large momentum come from small values of $R$, for
which the large $R$ behavior of the background is irrelevant. In
region I we have
\begin{equation}
\cala_N^{(I)}=\r_s^{-4}\ \lambda_N\sum_{\{n_i,m_i,s_i\}}\
\biggl(\prod_{i=1}^N\
 p_{2n_i+2s_i\nu_i,m_i} \biggr) \int_{R_{UV}}^{R_t}\ \frac{dR}{R}\
R^{\nb}\prod_{j=1}^N\ln^{m_j} (R),
\eqlabel{aikt}
\end{equation}
where
we again define
\begin{equation}
\nb\equiv -4+2N-\sum_{i=1}^N\nu_i+2\sum_{i=1}^N\nu_is_i
+2\sum_{i=1}^N n_i. \eqlabel{nbdefkt}
\end{equation}
Using (\ref{sumint}) and (\ref{sumnonint}), we note the following
properties of the coefficients $p_{a,b}$ which give the leading
large $Y$ contributions. For non-integer $\nu$
\begin{align}
\label{E:ptokappa1}
    p_{2n,n} &= C_{\nu} \rho_s^{2-\nu}(-b)^n Y^{2n} \kappa_{2n,0},\\
    p_{2\nu+2n,n} & = C_{\nu}
\rho_s^{2-\nu} Y^{2n+2\nu}(-b)^n (h_Y)^{\nu}
    \kappa_{2\nu+2n,0},
\end{align}
while for integer $\nu$
\begin{align}
    p_{2n,n}&= C_{\nu} \rho_s^{2-\nu}(-b)^{n} Y^{2n}\kappa_{2n,0}, & n<\nu\\
    p_{2n,n+1}&= C_{\nu} \rho_s^{2-\nu}(-b)^n Y^{2n}(\nu+1)^{-1}\kappa_{2n,1}, & n>\nu\\
    p_{2\nu+2s,s}&= C_{\nu} \rho_s^{2-\nu}(-1)^s (b Y^2)^{\nu+s}
        (\ln (Y))^{\nu+1}(\nu+1)^{-1}\kappa_{2\nu+2s,1}\\
    &=(-1)^{\nu}(\ln (Y))^{\nu+1}p_{2\nu+2s,\nu+s+1}.
\label{E:ptokappa6}
\end{align}

Now, we wish to evaluate the leading contributions to (\ref{aikt})
at large momentum, which come from the terms with the most
logarithmic contributions. Recalling that $0 \leq m_i \leq
n_i+\sigma_i(\nu_i+1)$ (where $\sigma_i\equiv s_i$ for $\nu_i$
integer and zero otherwise), leading logarithms in (\ref{aikt})
always come from factors related to the coefficients
(\ref{E:ptokappa1})-(\ref{E:ptokappa6}). Thus we can rewrite the
leading contributions to (\ref{aikt}) as
\begin{multline}
\cala_N^{(I)} \sim \r_s^{-4}\
    \lambda_N \sum_{\{n_i,s_i\}}\
    \\
    \times
    \left(\prod_{i=1}^N
    C_{\nu_i}
    \rho_s^{2-\nu_i}
    (-b)^{n_i+\nu_i \sigma_i}
    Y_i^{2 n_i+2 s_i\nu_i}
    (b \ln (Y_i))^{(s_i-\sigma_i) \nu_i}
    (\nu_i+1)^{-\sigma_i}
    \kappa_{2n_i+2s_i\nu_i,\sigma_i} \right)\\
    \times\ \int_{R_{UV}}^{R_t}\ \frac{dR}{R}\
    R^{\nb}(\ln (R))^{n_{tot}}
    \prod_{i=1}^N\left((-1)^{\nu_i} (\ln (Y_i))^{\nu_i+1}+(\ln
    (R))^{\nu_i+1}\right)^{\sigma_i}.
\end{multline}
To match with the earlier notation, we define $\mb \equiv \sum
\sigma_i$, and as in the AdS case we rearrange the indices so that
the first $\mb$ indices specify the $\sigma_i \neq 0$ contributions.
We need to evaluate the integral
\begin{equation}
    \int
        R^{\nb} (\ln (R))^{n_{tot}}
        \prod_{i=1}^{\mb}\left((-1)^{\nu_i} (\ln (Y_i))^{\nu_i+1}+(\ln
    (R))^{\nu_i+1}\right)
     \frac{dR}{R}.
\end{equation}

We start with the $\nb\neq 0$ case. It will be convenient to use the
following identity :
\begin{equation}
\label{E:modifiedA}
    \left((-1)^{\nu} (\ln (Y))^{\nu+1}+(\ln
    (R))^{\nu+1}\right) = \sum_{t=0}^{\nu} (-1)^t \begin{pmatrix} \nu+1 \\ t
        \end{pmatrix} (\ln (Y R))^{\nu+1-t} (\ln (Y))^t.
\end{equation}
The integral we wish to evaluate can be written as
\begin{equation}
    \int
        R^{\nb} (\ln (R))^{n_{tot}}
            \prod_{i=1}^{\mb}
        \left(
            \sum_{t=0}^{\nu_i} (-1)^t
            \begin{pmatrix}
                \nu_i+1 \\ t
            \end{pmatrix}
            (\ln ({Y_i} R))^{\nu_i+1-t} (\ln ({Y_i}))^t
        \right)
        \frac{dR}{R}.
\label{E:integral_1}
\end{equation}
In order to evaluate the leading term in (\ref{E:integral_1}), we
only need to keep track of the terms with the largest power of $(\ln
(Y_i))$ (the $t=\nu_i$ term). To see this, note that from
(\ref{E:logsintegral}) we find
\begin{multline}
    \int R^{\nb} (\ln (R))^{n_{tot}} (\ln ({Y}_1 R)) \ldots (\ln ({Y}_n R))
    \frac{d R}{R}
    \\
    =
    R^{\nb} (\ln (R))^{n_{tot}} \sum_{k=0}^n (-1)^k \frac{1}{{\nb}^{k+1}}
        \sum_{\substack{j_1 \ldots j_k \\ j_1 \neq \ldots \neq
        j_k}}^{n} \left( \prod_{i\neq j_1,\ldots j_k} \ln ({Y}_iR)
        \right)
  \\  +\mathcal{O}\left(
        R^{\nb} (\ln (R))^{n_{tot}-1} \prod \ln ({Y}_i R)
        \right).
\end{multline}
Thus,
\begin{multline}
    \int
        R^{\nb} (\ln (R))^{n_{tot}}
        \prod_{i=1}^{\mb}\left((-1)^{\nu_i} (\ln ({Y}_i))^{\nu_i+1}+(\ln
    (R))^{\nu_i+1}\right)
     \frac{dR}{R}
     \\
    =
    R^{\nb}(\ln (R))^{n_{tot}}
    \left(\prod_{i=1}^{\mb}(-1)^{\nu_i} (\nu_i+1)
        (\ln ({Y}_i))^{\nu_i}\right)
    \sum_{k=0}^{\mb} \frac{(-1)^k}{{\nb}^{k+1}}
        \sum_{\substack{j_1 \ldots j_k \\ j_1 \neq \ldots \neq
        j_k}}^{\mb} \left( \prod_{i\neq j_1,\ldots j_k} \ln ({Y}_iR)
        \right)\\
    +\mathcal{O}\left(R^{\nb}(\ln (R))^{n_{tot}-1} \prod_{i=1}^{\mb}(\ln ({Y_i}))^{\nu_i}
        \prod \ln ({Y}_i R) \right).
\end{multline}

This gives us a leading contribution of the form
\begin{multline}
\cala_N^{(I)} \sim \r_s^{-4}\
    \lambda_N \prod_{i=1}^N
    C_{\nu_i}\sum_{\{n_i,s_i\}}\
    \left(\prod_{i=1}^N
    \rho_s^{2-\nu_i}
    Y_i^{2 n_i+2 s_i\nu_i}
    (-b \ln (R))^{n_i}
(b \ln(Y_i))^{s_i \nu_i}
    \kappa_{2n_i+2s_i\nu_i,\sigma_i} \right)\\
    \times
    R^{\nb}
    \sum_{k=0}^{\mb}  \frac{(-1)^k}{{\nb}^{k+1}}
        \sum_{\substack{j_1 \ldots j_k \\ j_1 \neq \ldots \neq
        j_k}}^{\mb} \left( \prod_{i\neq j_1,\ldots j_k} \ln ({Y}_i
        R)
        \right)
    \Bigg|^{R_t}_{R_{UV}}
\end{multline}
which differs from its AdS counterpart (\ref{E:Aonebarnnonzero})
\begin{multline}
\cala_N^{(I)}=\r_s^{-4}\ \lambda_N \prod_{i=1}^N C_{\nu_i}
\sum_{\{n_i,m_i,s_i\}}\ \biggl(\prod_{i=1}^N\ \r_s^{2-\nu_i}
\ Y_i^{2n_i+2s_i\nu_i}\ \k_{2n_i+2s_i\nu_i,m_i} \biggr)\\
\times
    R^{\nb}
    \sum_{k=0}^{\mb} \frac{(-1)^k}{\nb^{k+1}}
    \sum_{\substack{j_1,\ldots,j_k \\ j_1 \neq\ldots\neq
    j_k}}^{\mb}\left( \prod_{i\neq j_1,\ldots,j_k} \ln (Y_i
    R)\right)\Bigg|^{R_t}_{R_{UV}}
\end{multline}
only by some powers of logs (as we found in the two-point
functions).

The analysis of the $\nb=0$ contributions in KT is again very
similar to that of AdS. Using (\ref{E:nbarzerointegral}) we find
that the $\nb=0$ term contributes
\begin{multline}
\eqlabel{nlb1}
\cala_N^{(I)} \sim \r_s^{-4}\
    \lambda_N \sum_{\{n_i,s_i\},\nb=0}\
    \sum_{t=0}^{n_{tot}+\sum_{i=1}^{\mb}(\nu_i+1)}\frac{(\ln(\mu R))^{t+1}}{(t+1)!}
    \partial^{(t)}P(-\ln(\mu))
    \Bigg|_{R_{UV}}^{R_t}
    \\
    \times
    \left(\prod_{i=1}^N
    C_{\nu_i}
    \rho_s^{2-\nu_i}
    (-b)^{n_i+\nu_i \sigma_i}
    Y_i^{2 n_i+2 s_i\nu_i}
    (b \ln (Y_i))^{(s_i-\sigma_i) \nu_i}
    (\nu_i+1)^{-\sigma_i}
    \kappa_{2n_i+2s_i\nu_i,\sigma_i} \right),
\end{multline}
where here
\begin{equation}
\label{E:defofPp0}
    P(x) = x^{n_{tot}}
    \prod_{i=1}^{\mb}\left((-1)^{\nu_i} (\ln
    (Y_i))^{\nu_i+1}+x^{\nu_i+1}\right).
\end{equation}
The AdS counterpart of this expression is given in
(\ref{E:Aonebarnzero}).

\subsubsection{The $R_{UV} \to 0$ limit}

We will now show that if the $\nu_i$ satisfy \eqref{E:casebnzero2},
which is the case whenever the $\lambda_N$ do not vanish (since we
are using the same couplings as we had in the AdS case and we assume
that the AdS case is renormalizable), then the divergences in
\eqref{aikt} coming from the $R_{UV}\to 0$ limit are non-analytic in
at most $(N-2)$ momenta, so that they correspond to contact terms.
Indeed, the first time a term non-analytic in $(N-1)$ momenta
appears as $R_{UV}\to 0$ is when $n_i=0$ for all $i$, and a single
$s_r=0$. In this case
\begin{equation}
{\bar n}=-4+2N+\sum_{i=1}^N\nu_i-2\nu_r\ge -4+2N+\nu_{tot}-2\nu_{max}.
\eqlabel{nbrdefkt}
\end{equation}
The right-hand side of \eqref{nbrdefkt} must be strictly positive
for $\lambda_N\ne 0$, so such a term is independent of $R_{UV}$ in
the $R_{UV}\to 0$ limit. Thus, whenever $\lambda_N$ is non-vanishing,
divergent terms in \eqref{aikt} as $R_{UV}\to 0$ are non-analytic in
at most $(N-2)$ momenta. This implies that
a cascading version of a conformal
field theory is holographically renormalizable whenever the original
conformal field theory is. By the same arguments as for the AdS
case, we also find that the $\prod C_{\nu_i}$ do not contribute to
non-analytic terms, and we will ignore them from here on.

\subsubsection{Analysis of leading terms}

As in the AdS case, we wish to consider $N$-point correlators which
are dominated by the region I contribution $\cala_N^{(I)}$.

Again, we introduce a typical momentum $k_*=Y_*/\rho_s$ with
\begin{equation}
Y_{\star}\equiv \frac{1}{N}\sum_{i=1}^N Y_i,
\end{equation}
and choose the separation between regions I and II to be at
\begin{equation}
R_t=\frac{1}{Y_*\ln^\gamma (Y_*)},\qquad \gamma> \frac 12,
\eqlabel{typmomckt}
\end{equation}
so that $R_t$ is in the overlap of regions I and II for all momenta,
provided they are not vastly different, $Y_i\sim Y_j$. By a
computation similar
to the AdS case, we find that
the region II contribution is dominated by\footnote{As in AdS, we
consider $\nu_{tot} > 2(N-2)$, as this is the only case where region
I will turn out to dominate over region II.}
\begin{equation}
    \cala^{II} \sim \lambda_N \rho_s^{2N-4-\nu_{tot}}
    R_t^{2N-4-\nu_{tot}}.
\end{equation}
Comparing
this to the region I contribution with
$\nb
> 0$, we find that region I dominates whenever
\begin{equation}
    \frac{(\gamma \ln(\ln (Y_{\star})))^{\mb}}
        {\ln (Y_{\star})^{(2\gamma-1)(\sum s_i \nu_i+\sum n_i)}} \gg 1
\end{equation}
which is always false, meaning that only terms with $\nb=0$ may
contribute if region I is to dominate over region II.
Since $\mu$ is independent of the momenta, we find that the
condition for the $\nb=0$ term in region I to dominate over region
II is
\begin{equation}
    n_{tot}+\sum_i s_i\nu_i+\bar{m}+1 > \gamma(4+\nu_{tot}-2N),
\end{equation}
 implying (using $\nb = 2 N - 4 - \nu_{tot} + 2 \sum \nu_i s_i + 2
 n_{tot} = 0$)
\begin{equation}
    \mb+1 > \left(\gamma-\frac{1}{2}\right)(4+\nu_{tot}-2N).
\label{case21kt}
\end{equation}
Again, this may always be satisfied for an appropriate choice of
$\gamma>\frac{1}{2}$.
The constraints we find
for the existence of the $\nb=0$ term are thus the same as those for
the AdS case, (\ref{E:casebnzero}) and
(\ref{E:casebnzero2}).

For the special
case of the three-point function with equal integer $\nu$,
we find that the leading term has $\mb=1$, so that (\ref{case21kt})
reduces to
\begin{equation}
\frac 12\ \frac{3\nu+2}{3\nu-2} > \gamma, \eqlabel{fincondition}
\end{equation}
which is consistent with $\gamma> \ft 12$.

\subsubsection{Leading expressions for correlation functions}

As in AdS, we may evaluate explicitly the correlation functions
which are dominated by the $\nb=0$ term in region I. These are
correlation functions for which there exists a choice of $n_i$ and
$s_i$ such that
\begin{equation}
    -4+2N+\nu_{tot} + 2(\sum_i s_i \nu_{i} - \nu_{tot}) + 2 n_{tot}=0.
\end{equation}
In this case, we find that to leading order in the large momentum
limit
\begin{multline}
\langle {\hat {\cal O}}_1(\vec{k}_1)\cdots
{\hat {\cal O}}_N(\vec{k}_N) \rangle
    \sim \delta(\sum \vec{k}_i)
    \lambda_N \sum_{\{n_i,s_i\in S\}}
    \left( \prod_{i=1}^N \kappa_{2n_i+2s_i\nu_i,\sigma_i}\right)\\
\times
    \left(\prod_{i=1}^N k_i^{2n_i+2s_i\nu_i}
        (b \ln (k_i/\Lambda))^{(s_i-\sigma_i)\nu_i}
        (-b)^{n_i+\nu_i \sigma_i}(\nu_i+1)^{-\sigma_i}
    \right)
    \\
    \times
    \sum_{t=0}^{n_{tot}+\sum_{i=1}^{\mb}(\nu_i+1)}
        \frac{(-\ln (k_{\star}/\Lambda))^{t+1}}{(t+1)!}
        \partial^{(t)} \biggl\{x^{n_{tot}} \prod_{
i=1
}^{\mb}
        \left((-1)^{\nu_i}(\ln (k_i / \Lambda)
        )^{\nu_i+1}+x^{\nu_i+1}\right)\biggr\}\Bigg|_{x=0},
\label{E:exactKT1}
\end{multline}
where we have set $\mu=1$ in order to write the solutions using
the natural scale $\Lambda=1/\rho_s$.
Notice that any apparent $\mu$-dependence in
expressions of the form (\ref{E:exactKT1}) is only through analytic
terms which disappear in position space.

To simplify this expression, we observe that if $t<n_{tot}$ the
second sum will vanish. Thus, (\ref{E:exactKT1}) can be rewritten as
\begin{multline}
\label{E:exactKT} \langle {\hat {\cal O}}_1(\vec{k}_1)\cdots
{\hat {\cal O}}_N(\vec{k}_N)
\rangle
    \sim \delta(\sum \vec{k}_i)
    \lambda_N \sum_{\{n_i,s_i \in S \}}
    \left( \prod_{i=1}^N \kappa_{2n_i+2s_i\nu_i,\sigma_i}\right)\\
\times    \left(\prod_{i=1}^N k_i^{2n_i+2s_i\nu_i} b^{s_i \nu_i}
        (\ln (k_i/\Lambda))^{(s_i-\sigma_i)\nu_i}
    \right)(b \ln (k_{\star}/\Lambda))^{n_{tot}}\ n_{tot}!
    \\
    \times
    \sum_{t=0}^{\sum_{i=1}^{\mb}(\nu_i+1)}
        \frac{(-\ln (k_{\star}/\Lambda))^{t+1}}{(n_{tot}+t+1)!}
        \partial^{(t)} \biggl\{\prod_{
i=1
}^{\mb}
        \left(\frac{(\ln (k_i/\Lambda)
        )^{\nu_i+1}+(-1)^{\nu_i}x^{\nu_i+1}}{\nu_i+1}\right)\biggr\}\Bigg|_{x=0}.
\end{multline}

For three-point functions we again find that there are two types of
contributions. Those from the (\ref{E:dominant1}) term are given
by
\begin{multline}
\langle {\hat {\cal O}}_1(\vec{k}_1)\cdots
{\hat {\cal O}}_3(\vec{k}_3) \rangle_a =
\delta(\vec{k}_1+\vec{k}_2+\vec{k}_3)
    \lambda_N
    \sum_{j=1}^3
    \sum_{\{n_i\} \in S_j}\\
    \times
    \left(\prod_{i=1}^3 (b k_i^{2}\ln (k_{\star}/\Lambda))^{n_i}\right)
    k_j^{2\nu_j} b^{\nu_j} \ n_{tot}!
    \left(\frac{-\ln (k_{\star}/\Lambda)
(\ln (k_j/\Lambda))^{\nu_j+1}}{(n_{tot} +1)!(\nu_j+1)}
     + \frac{(\ln (k_{\star}/\Lambda))^{\nu_j+2} \nu_j!}{(n_{tot} + \nu_j+2)!}
    \right)
    \\ \times
    \left( \prod_{i=1}^3
       \frac{(-1)^{n_i}\Gamma(\nu_i-n_i)}{2^{2n_i}\Gamma(n_i+1)\Gamma(\nu_i)}\right)
   \left(\frac{(-1)^{\nu_j}
        2}{\Gamma(1-\nu_j)\Gamma(\nu_j)}\right)^{\sigma_j}
    \left(\frac{-\Gamma(n_j-\nu_j+1)}{2^{2\nu_j}\Gamma(n_j+\nu_j+1)}\right),
\end{multline}
where we use the same notation as in (\ref{E:AdS3point}). The other
contributions come from the (\ref{E:dominant2}) term
\begin{multline}
\langle {\hat {\cal O}}_1(\vec{k}_1)\cdots
{\hat {\cal O}}_3(\vec{k}_3) \rangle_b =
\delta(\vec{k}_1+\vec{k}_2+\vec{k}_3)
    \lambda_N\\
    \times
    \sum_{j=1}^3
    \sum_{\{n_i\} \in S_j}
    \left(\prod_{i=1}^3 (b k_i^{2}\ln (k_{\star}/\Lambda))^{n_i}\right)
    \frac{(-\ln (k_{\star}/\Lambda))}{n_{tot}+1}
       \left( \prod_{i=1}^3
       \frac{(-1)^{n_i}\Gamma(\nu_i-n_i)}{2^{2n_i}\Gamma(n_i+1)\Gamma(\nu_i)}\right).
\end{multline}
The complete three-point function is given by
\begin{equation}
\langle {\hat {\cal O}}_1(\vec{k}_1)\cdots
{\hat {\cal O}}_3(\vec{k}_3) \rangle = \langle
{\hat {\cal O}}_1(\vec{k}_1)\cdots
{\hat {\cal O}}_3(\vec{k}_3) \rangle_a + \langle
{\hat {\cal O}}_1(\vec{k}_1)\cdots
{\hat {\cal O}}_3(\vec{k}_3) \rangle_b
\end{equation}
and should be understood in the same sense as the corresponding AdS
three-point correlation function \eqref{sumab}.

As a specific example, the three-point function for the massless
$\nu_i=2$ modes is given by
\begin{multline}
\langle {\hat {\cal O}}_2(\vec{k}_1)\cdots {\hat {\cal
O}}_2(\vec{k}_3) \rangle =
    \delta(\vec{k}_1+\vec{k}_2+\vec{k}_3)
    \lambda_N \times
    \\
    \frac{b^2}{48}
       \sum_{j=1}^3
    k_j^4
    \left((\ln (k_{\star}/\Lambda)) (\ln (k_j/\Lambda))^3-\frac{1}{4}(\ln
    (k_{\star}/\Lambda))^4\right).
   \end{multline}
As we did in the AdS case, we also consider the specific four-point
function with $\nu_1=\nu_2=5/2$ and $\nu_3=\nu_4=1/2$. We find that
the dominant contribution is of the form
\begin{multline}
    \frac{b}{12}\Bigg(
        k_1^2 (\ln (k_1/\Lambda))+k_2^2 (\ln (k_2/\Lambda)) -
        3 k_3^2 (\ln (k_3/\Lambda))- 3 k_4^2
        (\ln (k_4/\Lambda))
        \\
        -12 k_3 (\ln (k_3/\Lambda) )^{1/2}
            k_4 (\ln (k_4/\Lambda))^{1/2}
        \Bigg)
        \ln(k_{\star}/\Lambda)
\end{multline}
which may be compared to the exact AdS result (\ref{E:fourAdS}).

Unlike the AdS correlation functions which have a smooth $\nu_i\to
integer$ limit, correlation functions in asymptotically cascading
geometries do not have such a smooth limit in momentum
space.  This can be traced to the fact that unlike the AdS case, BtB
propagators in asymptotically cascading geometries do not have a
smooth $\nu_i\to integer$ limit in momentum space. However, since
the BtB propagator in position space does have a smooth limit (at
least for the leading terms which we computed), we believe that
higher order $N$-point correlation
functions in KT are smooth in $\nu$ in position space as well.

Finally, note that
the overall powers of momentum and of logarithms of momentum that we
find in the KT correlators are always given (at leading order) by
replacing $k \to k \sqrt{b \ln(k/\Lambda)}$ in the AdS correlators,
although the precise coefficients are different (as are the precise
momenta appearing in the logs, but this is something that we are not
sensitive to in our leading order computations).  This allows us to
easily verify that the normalized correlation functions indeed
depend on $N_{eff}$ as we expect.
In AdS,
dimensional arguments imply that a correlator $\vev{{\hat {\cal
O}}_{\nu_1} \cdots {\hat {\cal O}}_{\nu_N}}$ scales as
$k^{\nu_{tot}-2N+4}$, up to the overall delta function, and
sometimes up to logarithmic factors which disappear when we
transform to position space. This means that when we normalize the
correlation function by dividing by the norms (the square roots of
the two-point functions) of the operators, the correlator scales as
$k^{4-2N}$. According to the relation we found above between the KT
and AdS results, this implies that the normalized correlator in KT,
namely the correlator of the operators which we denoted by ${\tilde
{\cal O}}_{\nu}^{\prime}$ in \S3, scales as $k^{4-2N} (b
\ln(k/\Lambda))^{2-N}$ (up to the delta function of the momenta, and
sometimes up to additional logs which are the same in KT and in AdS
and which disappear when we Fourier transform to position space). We
expect normalized correlation functions in a large $N_{eff}$
$SU(N_{eff})$ gauge theory to scale as $N_{eff}^{2-N}$, so the
result we find is consistent with the identification $N_{eff}(k)
\propto b \ln(k/\Lambda)$. Of course, this is not surprising; in
standard gravity computations in AdS, the fact that all tree-level
correlators scale as $N^2$ (in some normalization) comes from the
fact that we can normalize the gravity action such that $N^2$ sits
in front of the action. Similarly, in the KT case we can normalize
the action so that $N_{eff}^2(\rho)$ sits in front of the gravity
action at the scale $\rho$, leading to the dependence on $N_{eff}$
that we found.

\section*{Acknowledgments}

We would like to thank Marcus Berg, Micha Berkooz, and Michael Haack
for valuable discussions. OA would like to thank the Aspen Center for
Physics and the Institute for Advanced Study for hospitality during
the course of this work. AB would like to thank the Aspen Center for
Physics and the Weizmann Institute
of Science for hospitality where part of this work was done. The work
of OA was supported in part by the Israel-U.S. Binational Science
Foundation, by the Israel Science Foundation (grant number 1399/04),
by the Braun-Roger-Siegl foundation, by the European network
HPRN-CT-2000-00122, by a grant from the G.I.F., the German-Israeli
Foundation for Scientific Research and Development, by Minerva, and by
a grant of DIP (H.52). AB's research at Perimeter Institute is
supported in part by the Government of Canada through NSERC and by the
Province of Ontario through MEDT. AB gratefully acknowledges support
by a NSERC Discovery grant.

\begin{appendix}
\section{Coefficients in the expansion of the KT BtB propagator}
\label{A:AppKTKKsolution}

We wish to find a perturbative solution for the BtB propagator of KK
modes in the asymptotically KT background \eqref{assKT}. Up to
integration constants and an overall power of $R^2$, the BtB
propagator is given by the solution to (\ref{E:KTKKEOM}),
\begin{equation}
    R^2 \psi^{\prime\prime} + R \psi^{\prime} -
    (\nu^2+Y^2 R^2\,h(R))\psi = 0.
\eqlabel{aeq1}
\end{equation}
We consider the case of  $\nu> 0$. The perturbative solution
crucially depends on whether $\nu$ is an integer or
not. We treat these two cases separately.

\subsection{Non-integer $\nu>0$}
\label{appnon}

To find a perturbative solution, we write the series expansion of
the field $\psi$ as either
\begin{equation}
    \psi(R) = R^{-\nu} \sum_{\substack{n=0 \\ m \le
    n}}^\infty p_{2n,m} R^{2n} \ln^m (R)
\eqlabel{sol1}
\end{equation}
or
\begin{equation}
    \psi(R) = R^{\nu} \sum_{\substack{n=0 \\ m \le
    n}}^\infty p_{2n+2\nu,m} R^{2n} \ln^m (R).
\eqlabel{sol2}
\end{equation}

Plugging \eqref{sol1} into the equation of motion \eqref{aeq1}, we find
\begin{multline}
    \sum p_{2n,m}R^{2n-\nu}\big((m-1)m \ln^{m-2}(R)+2
    m(2n-\nu)\ln^{m-1}(R)+4n(n- \nu)\ln^m(R)+\\
    R^2 Y^2(-a
    \ln^m (R)+ b \ln^{m+1} (R))\big)=0
\end{multline}
which may be rewritten as
\begin{equation}
\begin{split}
    &\sum_{n=0}^{\infty} \sum_{m=0}^{n} p_{2n,m}\ 4n(n -
     \nu)\  R^{2n-\nu} \ln^m (R)\\
    -&\sum_{n=1}^{\infty}\sum_{m=0}^{n-1} p_{2(n-1),m}\ Y^2
    a\ R^{2n-\nu} \ln^m (R)\\
    +&\sum_{n=1}^{\infty} \sum_{m=1}^{n} p_{2(n-1),m-1}\
    Y^2 b\ R^{2n-\nu} \ln^m (R)\\
    +&\sum_{n=1}^{\infty}\sum_{m=0}^{n-1}p_{2n,m+1}\
    2(m+1)(2n-\nu)\ R^{2n-\nu} \ln^m (R)\\
    +&\sum_{n=2}^{\infty} \sum_{m=0}^{n-2} p_{2n,m+2}\
    (m+2)(m+1)\ R^{2n-\nu} \ln^m (R)=0.
\end{split}
\label{E:serexpansion}
\end{equation}
We consider  first the leading $\ln^m (R)$ coefficients in the
series expansion \eqref{sol1}. From \eqref{E:serexpansion} we find
that $p_{0,0}$ is arbitrary and
\begin{equation}
\begin{split}
p_{2n,n}=&\frac{(-bY^2)}{4n(n-\nu)}\
p_{2(n-1),n-1}=\frac{(-bY^2)^{n}
\Gamma(1-\nu)}{2^{2n}\Gamma(n+1)\Gamma(n+1-\nu)}\ p_{0,0},\qquad
n\ge 1.
\end{split}
\eqlabel{negnu}
\end{equation}
Additionally, for any $n$ and $n> m\ge 0$
\begin{equation}
p_{2n,m}\propto p_{2n,n}\propto\left(bY^2\right)^n\ p_{0,0}.
\eqlabel{other}
\end{equation}

A similar analysis for \eqref{sol2} leads to arbitrary $p_{2\nu,0}$
and
\begin{equation}
\begin{split}
p_{2n+2\nu,n}=&\frac{(-bY^2)}{4n(n+\nu)}\ p_{2(n-1)+2\nu,n-1}=\frac{(-bY^2)^{n}\Gamma(1+\nu)}{2^{2n}\Gamma(n+1)\Gamma(n+1+\nu)}\ p_{2\nu,0},
\qquad n\ge 1
\end{split}
\eqlabel{negnu3}
\end{equation}
\begin{equation}
p_{2n+2\nu,m}\propto p_{2n+2\nu,n}\propto\left(bY^2\right)^n\ p_{2\nu,0}
\eqlabel{other2}
\end{equation}
for $n>m\ge 0$.

\subsection{Integer $\nu\ge 1$}
\label{appint}

To find a perturbative solution, we write the series expansion of
the field $\psi$ as
\begin{equation}
    \psi(R) = R^{-\nu} \biggl\{\sum_{n=0}^{\nu-1}\sum_{m=0}^n\  p_{2n,m} R^{2n} \ln^m (R)+
\sum_{n=\nu}^{\infty}\sum_{m=0}^{n+1}\  p_{2n,m} R^{2n} \ln^m (R)
\biggr\}. \eqlabel{sol3}
\end{equation}

Plugging \eqref{sol3} into the equation of motion \eqref{aeq1}, we
find
\begin{multline}
    \sum p_{2n,m}R^{2n-\nu}\big((m-1)m \ln^{m-2}(R)+2
    m(2n-\nu)\ln^{m-1}(R)+4n(n- \nu)\ln^m(R)+\\
    R^2 Y^2(-a
    \ln^m (R)+ b \ln^{m+1} (R))\big)=0,
\end{multline}
which may be rewritten as
\begin{equation}
\begin{split}
    &\sum_{n=0}^{\nu-1} \sum_{m=0}^{n} p_{2n,m}\ 4n(n -
     \nu)\  R^{2n-\nu} \ln^m (R)+\sum_{n=\nu}^{\infty} \sum_{m=0}^{n+1} p_{2n,m}\ 4n(n -
     \nu)\  R^{2n-\nu} \ln^m (R)\\
    +&\sum_{n=1}^{\nu}\sum_{m=0}^{n-1} -p_{2(n-1),m}\ Y^2
    a\ R^{2n-\nu} \ln^m (R)+\sum_{n=\nu+1}^{\infty}\sum_{m=0}^{n} -p_{2(n-1),m}\ Y^2
    a\ R^{2n-\nu} \ln^m (R)\\
    +&\sum_{n=1}^{\nu} \sum_{m=1}^{n} p_{2(n-1),m-1}\
    Y^2 b\ R^{2n-\nu} \ln^m (R)+\sum_{n=\nu+1}^{\infty} \sum_{m=1}^{n+1} p_{2(n-1),m-1}\
    Y^2 b\ R^{2n-\nu} \ln^m (R)\\
    +&\sum_{n=1}^{\nu-1}\sum_{m=0}^{n-1}p_{2n,m+1}\
    2(m+1)(2n-\nu)\ R^{2n-\nu} \ln^m (R) \\+&\sum_{n=\nu}^{\infty}\sum_{m=0}^{n}p_{2n,m+1}\
    2(m+1)(2n-\nu)\ R^{2n-\nu} \ln^m (R)\\
    +&\sum_{n=2}^{\nu-1} \sum_{m=0}^{n-2} p_{2n,m+2}\
    (m+2)(m+1)\ R^{2n-\nu} \ln^m (R)\\ +&\sum_{n=\nu}^{\infty} \sum_{m=0}^{n-1} p_{2n,m+2}\
    (m+2)(m+1)\ R^{2n-\nu} \ln^m (R)=0.
\end{split}
\eqlabel{Eserexpansion3}
\end{equation}
From \eqref{Eserexpansion3} we find that $p_{0,0}$ and $p_{2\nu,0}$
are arbitrary and \nxt For $1\le n<\nu$
\begin{equation}
\begin{split}
p_{2n,n}=&\frac{(-bY^2)}{4n(n-\nu)}\ p_{2(n-1),n-1}=\frac{(bY^2)^{n}
\Gamma(\nu-n)}{2^{2n}\Gamma(n+1)\Gamma(\nu)}\ p_{0,0},
\end{split}
\eqlabel{int1}
\end{equation}
with
\begin{equation}
p_{2n,m}\propto p_{2n,n}\propto\left(bY^2\right)^n\ p_{0,0}
\eqlabel{other2i}
\end{equation}
for $n>m\ge 0$. \nxt For $n=\nu$
\begin{equation}
\begin{split}
p_{2\nu,\nu+1}=&\frac{(-bY^2)}{2\nu(\nu+1)}\ p_{2(\nu-1),\nu-1}=
\frac{-(bY^2)^{\nu}}{2^{2\nu-1}\Gamma(\nu)\Gamma(\nu+2)}\ p_{0,0}.
\end{split}
\eqlabel{int2}
\end{equation}
\nxt For $n>\nu$
\begin{equation}
\begin{split}
p_{2n,n+1}=&\frac{(-bY^2)}{4n(n-\nu)}\ p_{2(n-1),n}=
\frac{(-bY^2)^{n-\nu}\Gamma(\nu+1)}{2^{2(n-\nu)}\Gamma(n-\nu+1)\Gamma(n+1)}\ p_{2\nu,\nu+1}\\
=&\frac{(-1)^{n-\nu+1}\nu
(bY^2)^n}{2^{2n-1}\Gamma(n+1)\Gamma(\nu+2)\Gamma(n-\nu+1)}\ p_{0,0}.
\end{split}
\eqlabel{int3}
\end{equation}

We also note from (\ref{Eserexpansion3}), that for a given $s$, the
highest power of $m$ for which an $R^{\nu+2s} \ln^m (R)$ term will
have a coefficient depending on $p_{2\nu,0}$ is at $m=s$, in which
case we find
\begin{equation}
\label{int4}
    p_{2\nu+2s,s} = C p_{0,0} + \frac{(-bY^2)^s\Gamma(\nu+1)}{2^{2s}
        \Gamma(s+1)\Gamma(s+\nu+1)} p_{2\nu,0},
\end{equation}
where $C$ is a constant which will not be important in our
calculations.

\section{Correlation functions when some $\nu_i$ approach integers}
\label{limit}
In this section we consider AdS correlation functions with index
$\nu_i = \hat{\nu}_i + \delta$ (for fixed integers $\hat{\nu}_i$),
in the $\delta \to 0$ limit. We will show explicitly that this
limit does not commute with the large momentum limit. That is, there
are terms in the correlation function which are not dominant at
large momentum when the $\nu_i$'s are non-integer, but that do
become dominant if we first take the $\nu_i \to \hat{\nu}_i$ limit.
Moreover, the leading non-analytic expression we extract for
integer-indexed correlation functions by this method agrees with our
direct method of calculation (\ref{E:exactAdS}).

We will explicitly discuss a specific limit of three-point
functions\footnote{Other limits can be analyzed similarly and lead
to the same conclusion.}. We consider a set of integer
$\hat{\nu}_i$'s and $\hat{n}_i$'s such that
\begin{equation}
2\sum_{i=1}\hat{n}_i=\sum_{i=1}^{3}\hat{\nu}_i-2\hat{\nu}_j-2
\eqlabel{apb1}.
\end{equation}
Following \eqref{E:exactAdS} and \eqref{E:AdS3point}, the leading
non-analytic contribution to the correlation function corresponding
to \eqref{apb1} takes the form
\begin{multline}
\langle \hat{\cal O}_1(\vec{k}_1)\cdots
\hat{\cal O}_3(\vec{k}_3) \rangle_{a}
=
    \lambda_3 \sum_{j=1}^3 \sum_{\{n_i \in S_j \}}
    \left(\prod_{i=1}^3 (b k_i^2)^{\hat{n}_i}\right)
    (b k_j^2)^{\hat{\nu}_j}
\\
    \times
    (-\ln (k_{\star}/\Lambda) )
    \left(\ln (k_j/\Lambda)-\frac{1}{2}(\ln
    (k_{\star}/\Lambda))\right)
    \left( \prod_{i=1}^3
        \kappa_{2\hat{n}_i,0}
    \right)
        \frac{\kappa_{2\hat{n}_j+2\hat{\nu}_j,1}}{\kappa_{2\hat{n}_j,0}}.
\label{apb2}
\end{multline}
We will concentrate only on the contribution of the $j$'th element
of the above sum.

Next, we consider a small deformation of $\hat{\nu}_j$ of the form
\begin{equation}
\hat{\nu}_j\to \nu_j=\hat{\nu}_j+\dd_j,\qquad \dd_j\ll 1.
\eqlabel{apb3}
\end{equation}
We would like to compare the $\dd_j \to 0$ limit of such a
correlation function to \eqref{apb2}. Clearly, \eqref{apb3} has no
$\bar{n}=0$ terms since it violates the condition \eqref{apb1}.
Therefore, region I does not dominate, and at finite $\delta_j$ one
can not evaluate the leading contribution to the correlation
functions by our methods. To obtain the leading contribution to the
($j$'th component of the) correlation function in the $\dd_j \to 0$
limit, we need to consider the $\bar{n}\propto\dd_j$ terms.
These terms will dominate in the $\dd_j \to
0$ limit.

We find that there are two important $\nb\propto \delta_j$ terms.
The first one is given by
\begin{equation}
\nb_{1}=-4+6-\sum_{i=1}^3\hat{\nu}_i-\dd_j+2\left(\hat{\nu}_j+\dd_j\right)+2\sum_{i=1}^3\hat{n}_i\\
    =\dd_j,
\eqlabel{apb4}
\end{equation}
where $\nb$ is the overall power of $R$, defined in \eqref{nbdef},
and in the last equality we used \eqref{apb1}. The second term comes
from the case where all the $s_i$'s are set to zero, but $\hat{n}_j$
is replaced by $n_j$
\begin{equation}
\hat{n}_j\to n_j=\hat{n}_j+\hat{\nu}_j
\eqlabel{apb5}
\end{equation}
resulting in
\begin{equation}
\nb_2=-4+6-\sum_{i=1}^3\hat{\nu}_i-\dd_j+2\left(\sum_{i=1}^3\hat{n}_i+\hat{\nu}_j\right)=-\dd_j.
\eqlabel{apb6}
\end{equation}
We will argue shortly that the other $\bar{n} \propto \dd_j$ terms
will be subdominant at large momentum.

The $\nb_1$ term contributes (see \eqref{ai})
\begin{equation}
\begin{split}
\cala^{(I)}_{3,1}=& \left(\prod_{i=1}^3 (b k_i^2)^{\hat{n}_i}\right)
(b k_j^2)^{\hat{\nu}_j} \rho_s^{-\delta_j} Y_j^{2\dd_j}
\left(\prod_{i=1}^3 \kappa_{2\hat{n}_i+2\cdot 0 \cdot
\hat{\nu}_i,0}\right)\ \frac{\kappa_{2\hat{n}_j+2(
\hat{\nu}_j+\dd_j),0}} {\kappa_{2\hat{n}_j+2\cdot 0 \cdot
\hat{\nu}_j,0}}\ \frac{(R_t)^{\dd_j}}{\dd_j},
\end{split}
\eqlabel{apb7}
\end{equation}
while the $\nb_2$ term contributes
\begin{equation}
\begin{split}
\cala^{(I)}_{3,2}=& \left(\prod_{i=1}^3 (b k_i^2)^{\hat{n}_i}\right)
(b k_j^2)^{\hat{\nu}_j} \rho_s^{-\delta_j} \left(\prod_{i=1}^3
\kappa_{2\hat{n}_i+2\cdot 0 \cdot \hat{\nu}_i,0}\right)\
\frac{\kappa_{2(\hat{n}_j+ \hat{\nu}_j)+2 \cdot 0\cdot
(\hat{\nu}_j+\dd_j),0}} {\kappa_{2\hat{n}_j+2\cdot 0 \cdot
\hat{\nu}_j,0}}\ \frac{(R_t)^{-\dd_j}}{(-\dd_j)}.
\end{split}
\eqlabel{apb8}
\end{equation}

We first note that taking
the $\delta \to 0$ limit results in
\begin{align}
    \kappa_{2n+2(\hat{\nu}+\delta),0}
        &\to \frac{1}{2\delta}
            \kappa_{2n+2\hat{\nu},1}+\mathcal{O}(1),
        &\\
    \kappa_{2n,0}
        &\to -\frac{1}{2\delta}
            \kappa_{2n+2\hat{\nu},1}+\mathcal{O}(1),
        &\hat{\nu}+\delta < n.
\end{align}
Therefore, the
non-analytic contributions to $\cala^{(I)}_{3,2}$ and
$\cala^{(I)}_{3,1}$ are of the form
\begin{align}
\cala^{(I)}_{3,1}= &\left(\prod_{i=1}^3 (b k_i^2)^{\hat{n}_i}\right)
    (b k_j^2)^{\hat{\nu}_j}
    \left(\prod_{i=1}^3 \kappa_{2\hat{n}_i,0}\right)
    \frac{\kappa_{2\hat{n}_j+2\hat{\nu}_j,1}}{2\delta_j\,\kappa_{2\hat{n}_j,0}}
    \\
    &\times \frac{1}{\dd_j}\left(-\ln (k_{\star}) \delta_j + \frac{1}{2}
        \left(-4\ln (k_j)\ln (k_{\star}) +\ln^2(k_{\star}) \right)\delta_j^2
        \right)+\mathcal{O}(\delta^{0} \ln (k_{*}),\delta),\\
    \cala^{(I)}_{3,2}=& \left(\prod_{i=1}^3 (b k_i^2)^{\hat{n}_i}\right)
        (b k_j^2)^{\hat{\nu}_j}
    \left(\prod_{i=1}^3 \kappa_{2\hat{n}_i,0}\right)
    \frac{-\kappa_{2\hat{n}_j+2\hat{\nu}_j,1}}{2\delta_j\, \kappa_{2\hat{n}_j,0}}
    \\
    &\times \frac{-1}{\dd_j}\left(\ln (k_{\star}) \delta_j + \frac{1}{2}
        \ln^2(k_{\star}) \delta_j^2
        \right)+\mathcal{O}(\delta^{0} \ln (k_{*}),\delta).
\eqlabel{E:apb9}
\end{align}
One can now easily take the $\delta_j \to 0$ limit to obtain
\eqref{apb2}. From this analysis, it is clear that the other
$\bar{n}\propto\delta_j$ terms will be of order $\delta_j^{-1}$, and
so will only contribute to order $\ln(k_{\star})$.

Similarly, it can be verified that the deformation \eqref{apb3} for
a set of integers satisfying
\begin{equation}
2\sum_{i=1}\hat{n}_i=\sum_{i=1}^{3}\hat{\nu}_i-2 \eqlabel{apb15}
\end{equation}
precisely reproduces in the $\dd_j\to 0$ limit the leading
non-analytic contribution \eqref{E:AdS3point2}.

\end{appendix}

\end{document}